\documentclass[aps,prx,twocolumn,amsmath,amssymb,10pt,a4paper,groupedaddress,superscriptaddress]{revtex4-1}
\pdfoutput=1

\usepackage{hyperref}

\hypersetup{
        colorlinks=true,
        linktoc=all,
        linkcolor=violet,
        urlcolor=blue
}

\usepackage{amsmath,amssymb}
\usepackage{geometry}
\usepackage{amsmath, amssymb}
\usepackage{graphicx}
\usepackage{bbold}
\usepackage{algorithm}
\usepackage{algpseudocode}
\usepackage{amsthm}
\usepackage{mathtools}
\usepackage{MnSymbol}

\newcommand{\be}{\begin{equation}}
\newcommand{\ee}{\end{equation}}
\newcommand{\bd}{\begin{displaymath}}
\newcommand{\ed}{\end{displaymath}}
\newcommand{\BE}{\begin{eqnarray}}
\newcommand{\EE}{\end{eqnarray}}

\newcommand{\bra}{\left\langle}
\newcommand{\ket}{\right\rangle}

\newcommand{\id}{{\rm 1\!\!I}}

\newcommand{\bj}{\ensuremath{\mathbf{j}}}

\newcommand{\bs}{\ensuremath{\mathbf{s}}}

\newcommand{\tr}{\mathrm{Tr}}

\newcommand{\mcM}{\mathcal{M}}

\newcommand{\mcA}{\mathcal{A}}

\newcommand{\mcS}{\mathcal{S}}

\graphicspath{{../../figs/}}

\usepackage{epigraph,varwidth}
\usepackage{etoolbox}
\usepackage{lipsum}

\usepackage{tcolorbox}
\newtcolorbox{mybox}{colback=green!5!white,colframe=green!75!black}

\usepackage[normalem]{ulem}

\begin{document}
\title{Quantum-inspired memory-enhanced stochastic algorithms}
\author{John Realpe-G\'omez}\email{john.realpe@gmail.com} 
\affiliation{Instituto de Matem\'aticas Aplicadas, Universidad de Cartagena, Bol\'ivar 130001, Colombia}
\author{Nathan Killoran}
\affiliation{Xanadu, Toronto, Canada}

\begin{abstract}
Stochastic models are highly relevant tools in science, engineering, and society. Recent work suggests emerging quantum computing technologies can substantially decrease the memory requirements for simulating stochastic models. Here we show that some of these recent quantum memory-enhanced algorithms can be either implemented or approximated classically. 
In other words, we show that it is possible to develop quantum-inspired classical algorithms that require much less memory than the best classical algorithms known to date. Being classical, such algorithms could be implemented in state-of-the-art high-performance computers, which could potentially enhance the study of large-scale complex systems. Furthermore, since memory is the main bottleneck limiting the performance of classical supercomputers in one of the most promising avenues to demonstrate quantum `supremacy', we expect adaptations of these ideas may potentially further raise the bar for near-term quantum computers to reach such a milestone. 
\end{abstract}

\maketitle

\section{Introduction}

From the prediction and understanding of financial markets~\cite{bouchaud2018trades,bisias2012survey,farmer2009economy} or the intricate relationships among ecosystems' resilience, climate change, and human activity~\cite{stern2016economics,cai2015environmental,franzke2015stochastic}, to the design and operation of the artificial intelligence architectures~\cite{lake2015human,ghahramani2015probabilistic,goodfellow2016deep,murphy2012machine} that pervade our lives, stochastic models and the corresponding algorithms to implement them are indispensable. Optimizing the computational resources, such as speed and memory, needed for running stochastic algorithms like Markov chain Monte Carlo, is crucial for keeping up with the fast-growing and increasingly complex problems our society faces.

With the emerging commercialization of quantum computing technologies and the race to demonstrate quantum `supremacy'~\cite{mohseni2017commercialize, boixo2018characterizing}, there is much interest in understanding what aspects of information processing quantum computers can do better than their classical counterparts. 
While much work has focused on potential quantum speedup~\cite{biswas2017nasa}, there is growing interest in potentially extreme memory reductions that quantum protocols can provide~\cite{gu2012quantum,thompson2018causal,aghamohammadi2018extreme,ghafari2018single,ghafari2019interfering,palsson2017experimentally,binder2018practical,liu2018optimal,thompson2017using}.  
Indeed, both theoretical~\cite{gu2012quantum,thompson2018causal,aghamohammadi2018extreme,binder2018practical,liu2018optimal,thompson2017using} and experimental~\cite{ghafari2018single,ghafari2019interfering,palsson2017experimentally} work suggests that quantum protocols can generate samples from stochastic processes using much less memory than the best classical counterparts known to date, i.e., the so-called (classical) $\epsilon$-machines~\cite{crutchfield1989inferring,shalizi2001computational}. These $\epsilon$-machines rely on the minimum `deterministic' information about the past of a stochastic process necessary to statistically predict its future---known as \emph{causal states}.  

For instance, imagine a coin inside a box that is regularly perturbed. Each perturbation of the box can flip the coin with a probability $p$, irrespective of the current state of the coin (see Fig.~\ref{f:qis_fig1}). If $p\neq\tfrac{1}{2}$, all we need to know is the previous state of the coin, i.e., the two causal states \emph{heads} or \emph{tails}, to correctly predict the probabilities of all future trajectories; therefore, we need to save in memory one bit of information. The information saved is `deterministic' in the sense that it is either heads or tails and not a classical probabilistic mixture of the two. In contrast, quantum $\epsilon$-machines can operate using less memory by encoding information about the past on quantum superpositions, the so-called {\em quantum} causal states (see Fig.~\ref{f:qis_fig2}). If $p=\tfrac{1}{2}$ the process is equivalent to tossing a coin at random at each moment, so the amount of memory required by a classical $\epsilon$-machine discontinuously jumps to zero. The amount of memory required by the quantum $\epsilon$-machine also goes to zero, though continuously, so it cannot do better in this case (see Fig.~\ref{f:memory}).

Such quantum-enabled memory reductions are reflected in two possible ways. First, in the sequential generation of a single stochastic trajectory due to a smaller state space of the quantum $\epsilon$-machine, e.g., an $\epsilon$-machine that needs a trit of memory could be simulated by a quantum $\epsilon$-machine that needs to keep in memory only a qubit~\cite{thompson2018causal, ghafari2018single}---this is referred to as topological memory reduction. Second, in the parallel generation of a large number of stochastic trajectories, or samples, due to a smaller (von Neumann) entropy of the quantum $\epsilon$-machine at the stationary state~\cite{gu2012quantum,ghafari2019interfering,palsson2017experimentally}---this is referred to as statistical memory reduction. The quantum-enabled reduction of topological memory relies on finding a representation of the quantum causal states in terms of a quantum system of smaller dimension. In contrast, following the Schumacher quantum coding theorem~\cite{schumacher1995quantum}, the quantum-enabled reduction of statistical memory relies on finding a suitable quantum coding of a large number of sample quantum causal states, distributed according to the stationary state of the stochastic process.

Except for a very recent work~\cite{liu2018optimal}, all quantum $\epsilon$-machines studied to date encode information only on the amplitude of quantum states, i.e., no phase information is used~\cite{aghamohammadi2018extreme,ghafari2018single,ghafari2019interfering,palsson2017experimentally, thompson2018causal, binder2018practical,gu2012quantum}. Such `amplitude-encoded' quantum $\epsilon$-machines, however, can lead to `extreme' memory reductions~\cite{aghamohammadi2018extreme}, i.e., to situations where the memory required by the classical $\epsilon$-machine diverges while that required by the corresponding quantum $\epsilon$-machine remains finite. 

But, if the information encoded in a probability distribution and that encoded in its square root is the same, why is there a difference in the memory requirements of classical and quantum $\epsilon$-machines? Indeed, an early algorithm for `quantum deep learning'~\cite{wiebe2014quantum} that only exploited the amplitude of quantum states was later shown to be implementable classically~\cite{wiebe2015quantum}. Nowadays, there is growing interest in such `quantum-inspired' algorithms~\cite{tang2018quantum,gilyen2018quantum,hen2018quantum,arrazola2019quantum}. This aligns with recent work~\cite{realpe2017modeling,realpe2018cognitive} suggesting that classical message-passing algorithms can be written in a way mathematically analogous to quantum dynamics in imaginary-time, i.e., by changing time $t$ into $- i t$ in the Schr\"odinger equation. In particular, square roots of probabilities arise naturally in such a classical setting. But when the phase degree of freedom is never used, the imaginary unit does not play any role, and it is natural to expect that both real-time and imaginary-time quantum dynamics can become similar.

Here we provide evidence that in some cases it is possible to classically implement such `amplitude-encoded' quantum $\epsilon$-machines. More precisely, we show that it is sometimes possible to implement classical protocols that have the same memory reductions as amplitude-encoded quantum $\epsilon$-machines. We do so by saving information about the past of a stochastic process using probabilistic mixtures, or {\em stochastic} causal states, rather than the deterministic causal states of standard $\epsilon$-machines. We also show that it is sometimes possible to build classical algorithms that, by exploiting an operational interpretation of negative numbers arising in a decomposition of probability distributions, introduced in Refs.~\cite{realpe2017modeling,realpe2018cognitive}, can generate stochastic trajectories while using much less memory than the corresponding classical $\epsilon$-machines. Interestingly, in the case of a coin being regularly flipped with probability $p$ mentioned above, the memory requirements approach zero continuously as $p\to \tfrac{1}{2}$, much as the quantum $\epsilon$-machine does, instead of jumping discontinuously to zero as the classical $\epsilon$-machine does. Such algorithms, though, do not attain the same memory reductions of amplitude-encoded quantum $\epsilon$-machines, so there is room for improvement. In particular, they do not exploit square roots of probabilities. We discuss how belief propagation can pave the way for expoiting such square roots. 

Our results therefore show that it is possible to develop quantum-inspired classical stochastic algorithms that require much less memory than the best classical stochastic algorithms known to date. Now, memory is the main bottleneck limiting the performance of classical supercomputers in one of the most promising avenues to demonstrate quantum `supremacy' in the near term~\cite{boixo2018characterizing}. So we expect adaptations of these ideas may potentially further raise the bar for quantum computing technologies to reach such a milestone.

The rest of this paper is organized as follows. In Sec.~\ref{s:frame} we review the main concepts related to $\epsilon$-machines (Sec.~\ref{s:classical}) as well as their quantum counterparts (Sec.~\ref{s:quantum}), and present two specific examples that have been recently demonstrated experimentally~\cite{ghafari2019interfering,palsson2017experimentally,ghafari2018single}, which we will revisit in the next sections.  In Sec.~\ref{s:results} we identify some features implicit in quantum $\epsilon$-machines and discuss how these can be implemented classically. Based on these, we introduce in Sec.~\ref{s:c-examples} two classical algorithms for the two examples mentioned above that require less memory to operate than the best algorithms known to date, i.e., their corresponding $\epsilon$-machines. One of these algorithms reaches the same memory gains as the corresponding quantum $\epsilon$-machine, while the other does not. In Sec.~\ref{s:extensions} we discuss the extension of the main ideas developed previously to build general quantum-inspired memory-enhanced algorithms. In particular, in Sec.~\ref{s:gqi} we introduce a general quantum-inspired classical algorithm that can generate samples with much less statistical memory than the best classical algorithms known to date. However, this algorithm does not reach, in general, the same memory savings of the corresponding quantum protocols. In Appendix~\ref{s:exqi} we present a detailed example. In Sec.~\ref{s:sqrt_bp} we briefly discuss how recent work~\cite{realpe2017modeling, realpe2018cognitive}, showing that belief propagation on chain- and cycle-like graphical models can be formulated as quantum-like dynamics, could serve as a general framework to exploit square roots of probabilities. In Appendix~\ref{s:bp} we present a detailed discussion of these ideas and introduce two graphical models whose dynamics is mathematically analogous to the two quantum $\epsilon$-machines discussed as examples in Sec.~\ref{s:quantum}. Finally, in Sec.~\ref{s:conclusions} we present the conclusions of this work and put the ideas introduced here in a broader perspective.

\section{Framework}\label{s:frame}
\subsection{Classical stochastic algorithms}\label{s:classical}
\begin{figure}
\centering
\includegraphics[width=\columnwidth]{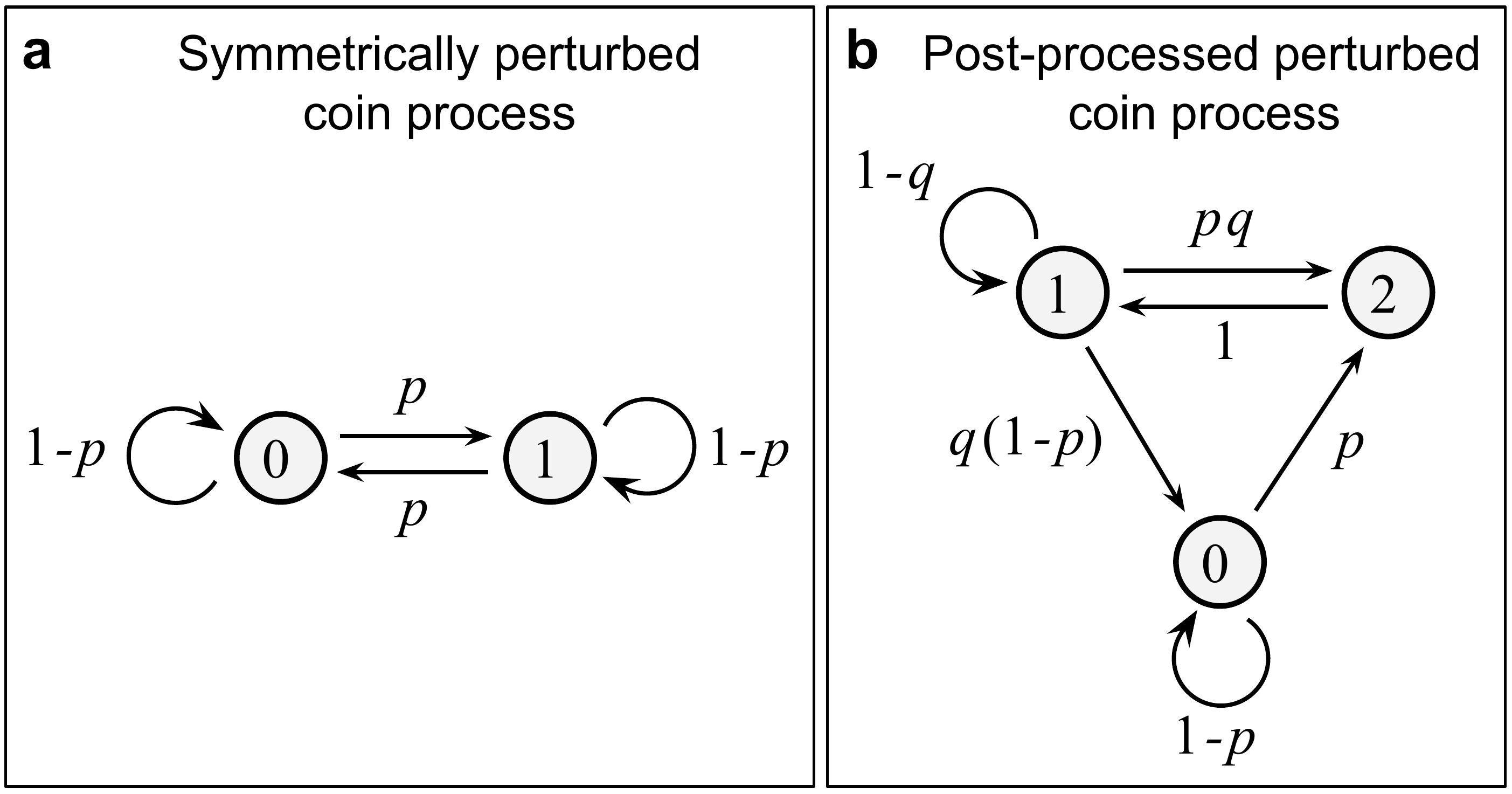}
\caption{{\em Some stochastic models studied here:} (a) Symmetrically perturbed coin process studied in Refs.~\cite{gu2012quantum,ghafari2019interfering, palsson2017experimentally}. It consists of a coin that is regularly flipped with probability $p$ (horizontal arrows), independently of whether the state of the coin (circles) is heads (0) or tails (1). With probability $1-p$ the coin remains in the same state (circular arrows). (b) Post-processed perturbed coin process analyzed in Refs.~\cite{thompson2018causal,ghafari2018single}. Consider a coin that is flipped regularly from heads (0) to tails (1) with probability $p$ and vice versa with probability $q$. With probabilities $1-p$ and $1-q$, respectively, the coin remains in heads or tails. The process illustrated in (b) is obtained by replacing the last 0 in each consecutive substring of 0s with a 2 (for example, $\cdots 00001101001\cdots$ becomes $\cdots 00021121021\cdots$). So, 0 can be followed by 0 or 2, while 1 can be followed by anything, and 2 can only be followed by 1. The probabilities for each transition are displayed beside the arrows (see Ref.~\cite{thompson2018causal} for further details).
}\label{f:qis_fig1}
\end{figure}
Consider a system whose state at time step $t\in\mathbb{Z}$ can be described by a stochastic variable $X_t$ which takes a value $x_t\in \mathcal{A}$ in a certain alphabet $\mathcal{A}$. The system's dynamics can then be described by a sequence of stochastic variables $\overset{\leftrightarrow}{X} = \overset{\leftarrow}{X}\overset{\rightarrow}{X}$. Here $\overset{\leftarrow}{X}=\dotsc X_{-2} X_{-1}$ and $\overset{\rightarrow}{X}=X_0X_1\dotsc$  are the sequences describing, respectively, the past and future dynamics of the system. The system's dynamical law is specified by a probability distribution $P(\overset{\leftarrow}{X},\overset{\rightarrow}{X})$. Sampling from a given stochastic process amounts to generating a sequence of variables from the corresponding dynamical law $P(\overset{\leftarrow}{X},\overset{\rightarrow}{X})$. A na\"ive way to sample a future sequence $\overset{\rightarrow}{X}$, given a realization of the past sequence $\overset{\leftarrow}{x}$, may require infinite memory; for instance, if we need to store the whole sequence $\overset{\leftarrow}{x}$.

More compact representations of stochastic processes that require less memory resources are therefore highly desirable. Sometimes the probability distribution $P(\overset{\leftarrow}{X},\overset{\rightarrow}{X})$ can be factorized into simpler probability distributions. For instance, in the case of stationary $m$th-order Markov chains, the whole dynamics can be generated by a single conditional probability distribution $p(x_{t}|x_{t-m},\dotsc , x_{t-1})$ that yields the probability that the state of the system at time step $t$ is $x_t$, given that the previous $m$ states the system visited were $x_{t-m},\dotsc, x_{t-1}$. A very common example with $m=1$ is the Markov chain Monte Carlo algorithm. 

However, in some cases the order of the Markov chain, $m$, can be prohibitively large or even infinity (see e.g., Fig. 1 in Ref.~\cite{thompson2018causal}). Fortunately, such long-range temporal correlations can sometimes be more compactly captured via hidden variables. A hidden Markov model (HMM) is characterized by a set of hidden variables or states $\mathcal{S}$, an alphabet $\mathcal{A}$, and the probability $P(x,j|i)$ that if the HMM is in state $i\in\mathcal{S}$ it emits output $x\in\mathcal{A}$ and transitions to state $j\in\mathcal{S}$ (see Fig.~\ref{f:qis_fig1}). 

The most direct way to characterize the memory requirements of a HMM is perhaps by the dimension of its state space. The {\em topological memory}
\be\label{e:Dc}
D_c = \log_2|\mathcal{S}|,
\ee
corresponds to the number of bits necessary to encode the $|\mathcal{S}|$ states in which a HMM can be. The topological memory characterizes the memory required for sequential sampling, i.e., for generating a single sample trajectory. 

A perhaps more subtle way to characterize the memory requirements of a HMM is by the amount of information actually encoded in the states of a HMM, once it has reached the stationary state $\pi$. The {\em statistical memory}
\be\label{e:Hc}
H_c = -\sum_{j\in\mathcal{S}}\pi_j\log_2\pi_j,
\ee
corresponds to the entropy of the HMM's stationary sate $\pi$. The statistical memory characterizes the memory required for parallel sampling, i.e., for the simultaneous generation of $M\gg 1$ samples at the stationary state. This observation is based on Shannon's source coding theorem which states that $M\gg 1$ samples distributed according to $\pi$ can be encoded into about $M H_c$ bits. The statistical memory of a HMM with state space $\mathcal{S}$ can be considered as the topological memory of a new HMM with state space $\mathcal{S}^M$. Such a new HMM operates at the stationary state and transitions $\bj\to \bj^\prime$, with $\bj ,\bj^\prime\in\mathcal{S}^M$, correspond to the independent transitions $j_\ell\to j_\ell^\prime$, with $\ell = 1\dotsc , M$, of the $M$ samples of the original HMM. According to Shannon source coding theorem, at the stationary state $\pi$ we can then find an encoding of the state space $\mathcal{S}^M$ such that the set of states $\mathcal{S}^\ast$ defined on the code space is of size $\sim 2^{M H_c}$.

The best HMMs known to date, in the sense that they require the smallest topological and statistical memory, are called $\epsilon$-machines. The states of an $\epsilon$-machine are given by a mapping $\epsilon$ that encodes an equivalence relation 
\be\label{e:causal}
\overset{\leftarrow}{x}\sim_\epsilon\overset{\leftarrow}{y} \Leftrightarrow P(\overset{\rightarrow}{X}|\overset{\leftarrow}{x}) = P(\overset{\rightarrow}{X}|{\overset{\leftarrow}{y}}),
\ee
with $\epsilon(x) = \epsilon(y) = j\in\mathcal{C}$, where $\mathcal{C}$ is the set of states of the $\epsilon$-machine, which we will refer to as (deterministic) {\em causal states}. An important property of any $\epsilon$-machine, called unifilarity and implied by Eq.~\eqref{e:causal}, is that the causal state $j$ to which it transitions at any given time is completely determined by the output $x$ emitted in such a transition and the state $i$ from which the transition takes place. More precisely,
\be\label{e:unifil}
P(x,j|i)= P(x|i)\delta_{j, f(i,x)},
\ee
where $\delta_{j, k}$ is the Kronecker delta function and ${f: \mathcal{S}\times\mathcal{A}\to\mathcal{S}}$ is a deterministic function that returns the causal state $f(i,x)\in\mcS$ to which the HMM transitions from state $i\in\mcS$ when it emits output $x\in\mcA$.

In general, however, there is room for improvement since $\epsilon$-machines do not always reach the minimum memory requirement, which is given by the mutual information between past and future ${I(\overset{\leftarrow}{X}:\overset{\rightarrow}{X})}$. We now know~\cite{gu2012quantum,binder2018practical,liu2018optimal,aghamohammadi2018extreme,ghafari2018single,thompson2018causal} quntum models can do better in such cases. 
\subsection{Quantum-enhanced stochastic algorithms}\label{s:quantum}
\subsubsection{General considerations}
We now discuss recent work~\cite{gu2012quantum,binder2018practical,liu2018optimal,aghamohammadi2018extreme,ghafari2018single,thompson2018causal} on quantum protocols for generating samples from a given classical $\epsilon$-machine that can operate with less memory requirements, the so-called quantum $\epsilon$-machines. The central idea is to define quantum causal states as
\be\label{e:amplitude}
\left|\xi_i\ket = \sum_{x\in\mcA , j\in\mcS}\sqrt{P(x,j|i)} \left|j\ket\left| x\ket ,
\ee
where $\{\left|j\ket\}_{j\in\mcS}$ is an orthonormal basis of a quantum system that represents the deterministic causal states $j\in\mcS$ of the classical $\epsilon$-machine, and $\{\left|x\ket\}_{x\in
\mcA}$ is an orthonormal basis of another quantum system that represents the corresponding outputs $x\in\mcA$. 

It is always possible~\cite{binder2018practical} to devise a unitary quantum operator $U$, such that
\be\label{e:U}
U\left|\xi_i\ket\left|0\ket = \sum_{x\in\mcA}\sqrt{P(x|i)}\left|\xi_{f(i,x)}\ket\left|x\ket.
\ee
If we measure the second system in Eq.~\eqref{e:U} in the basis $\{\left|x\ket\}_{x\in\mcA}$, that system will output $x\in\mcA$ with the correct probability $P(x|i)$ and the first system in Eq.~\eqref{e:U} will transition to the correct next quantum causal state $\left|\xi_{j^\ast}\ket$, with $j^\ast={f(i,x)}$, so the protocol can be applied iteratively. Importantly, except for a very recent work~\cite{liu2018optimal} which shows that adding phases to Eq.~\eqref{e:amplitude} can provide further memory reductions, all work to date has only considered amplitude-encoded quantum causal states like those in Eq.~\eqref{e:amplitude}. In this work we will exclusively focus on the latter which, while being less general, can still provide extreme memory reductions, i.e., situations where the memory required for a classical $\epsilon$-machine to operate diverges while that required by the corresponding amplitude-encoded quantum $\epsilon$-machine remains finite~\cite{aghamohammadi2018extreme}.

At the stationary state $\pi$, the quantum causal state $\left|\xi_i\ket$ appears with probability $\pi_i$. So the state of the quantum system can be represented by a density matrix
\be\label{e:rho1}
\rho=\sum_{i\in\mcS}\pi_i\left|\xi_i\ket\bra\xi_i\right|,
\ee
whose rank yields the {\em quantum topological memory}
\be
D_q = \log_2\left[\textrm{rank}(\rho)\right],
\ee
and whose von Neuman entropy yields the {\em quantum statistical memory}
\be\label{e:qstat}
S_q = -\tr\rho\log_2\rho .
\ee

Unlike deterministic casual states $\{\left|j\ket_A\}_{j\in\mcS}$ that are always orthogonal, the quantum causal states introduced in Eq.~\eqref{e:amplitude} can be non-orthogonal, i.e., we can have $\bra\xi_i|\xi_j\ket\neq 0$ for $i\neq j$ and therefore $\rho$ can be non-diagonal. So the quantum statistical memory $S_q$ can be strictly lower than the classical one $H_c$. 
This implies that if we want to generate $M$ independent samples in parallel we can encode the corresponding density matrix $\bigotimes_{m=1}^M \rho$ with about $M S(\rho) \leq M H_c$ qubits describing the typical subspace~\cite{schumacher1995quantum,nielsen2002quantum}. 

\subsubsection{Examples}\label{s:qex}
\begin{figure*}
\centering
\includegraphics[width=\textwidth]{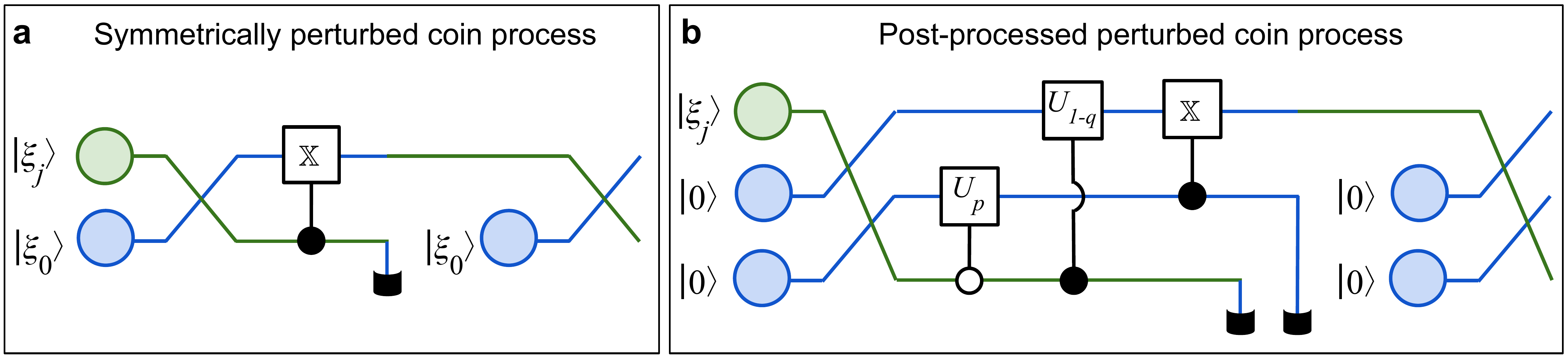}
\caption{{\em Quantum protocols studied here:} (a) Quantum $\epsilon$-machine for simulating the Markov chain in Fig.~\ref{f:qis_fig1}a. The protocol starts with the current causal state $\left|\xi_j\ket$ (see Eqs.~\eqref{e:S1}-\eqref{e:S2}) and an ancillary qubit in state $\left|\xi_0\ket$. It then applies to the combined system a controlled \texttt{NOT} gate (see Eq.~\eqref{e:CNOT2}) where the first qubit is the control. Finally, the first qubit is measured, producing the corresponding output with the correct probabilities and the second qubit transitions to the correct next causal state, so the protocol can be iterated. (b) Quantum $\epsilon$-machine for simulating the Markov chain in Fig.~\ref{f:qis_fig1}b. The protocol starts with the current causal state $\left|\xi_j\ket$ (see Eqs.~\eqref{e:|0'>}-\eqref{e:|2'>}) and two ancillary qubits, both in state $\left|0\ket$. It then applies a negated controlled unitary $U_p$ (see Eq.~\eqref{e:-CUp}) between the first and third qubits, where the first qubit is the control. Next, it applies a controlled unitary $U_{1-q}$ (see Eq~\eqref{e:CUq})  between the first and second qubits, where the first qubit is the control. It finally applies a controlled \texttt{NOT} gate (see Eq.~\eqref{e:CNOT}) between the second and third qubits, where the third qubit is the control. Afterwards, the first and third qubits are measured giving outputs $y_1$ and $y_3$, respectively; the probability to have both outputs equal to one is zero. Therefore, the variable $x = y_1+2\, y_3\in\{0,1,2\}$ yields the outputs corresponding to the Markov chain in Fig.~\ref{f:qis_fig1}b with the correct probabilities, and the second qubit transitions to the correct next quantum causal state, so the protocol can be iterated. }\label{f:qis_fig2}
\end{figure*}
\begin{figure}
\centering
\includegraphics[width=\columnwidth]{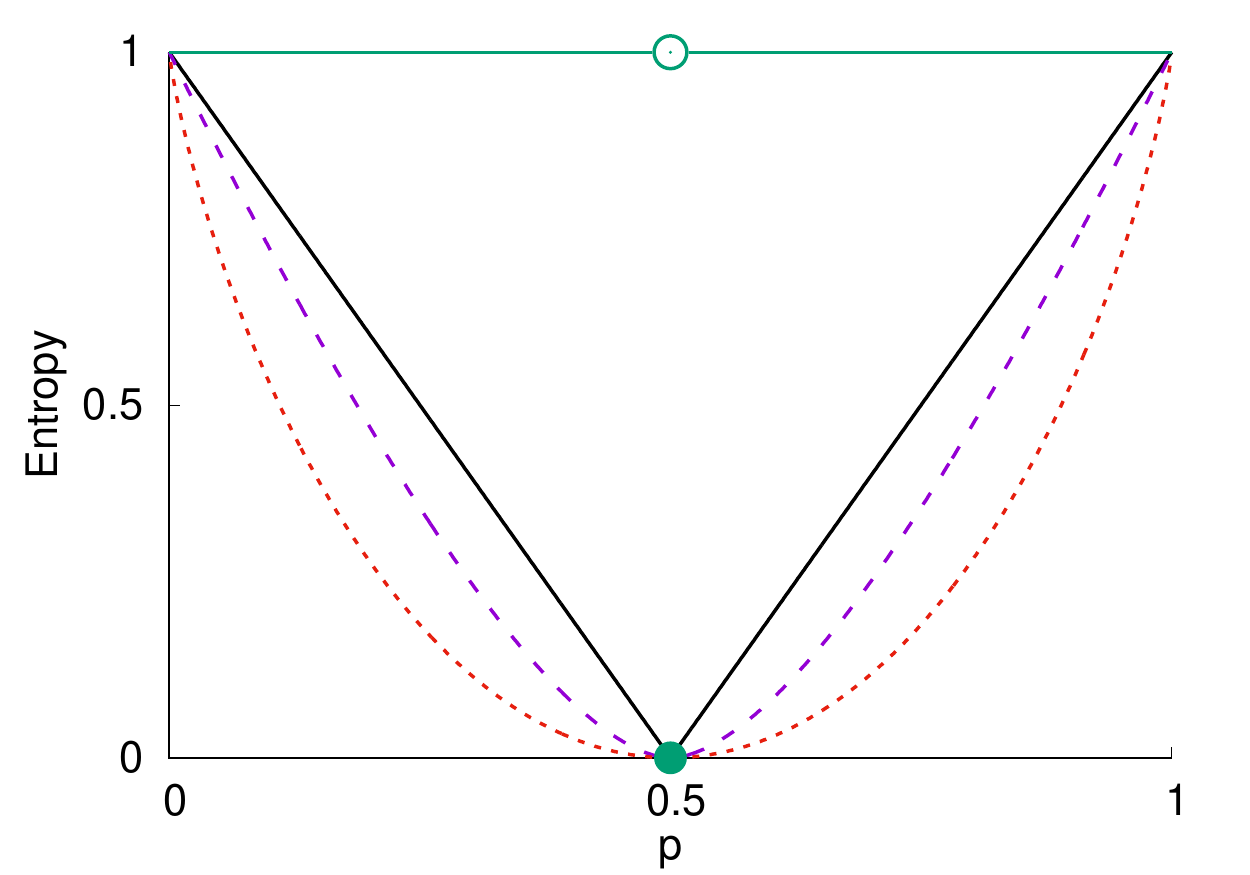}
\caption{{\em Memory requirements of different algorithms:} The best classical algorithm known to date to simulate the symmetrically perturbed coin process (see Fig.~\ref{f:qis_fig1}a), i.e., its classical $\epsilon$-machine, requires to keep in memory one bit when $p\neq \tfrac{1}{2}$ (green horizontal straight line) and zero bits when $p=\tfrac{1}{2}$ (filled green circle). The dashed purple curve shows the memory requirements for the corresponding amplitude-encoded quantum $\epsilon$-machine. They are substantially smaller and reach zero continuously as $p$ approaches $\tfrac{1}{2}$. The black triangular solid curve shows the $|1-2p|$ bits per sample the classical algorithm introduced here (see Fig.~\ref{f:qis_fig4}) needs to keep in memory. Interestingly, the number of bits needed also goes to zero continuously as $p$ approaches $\tfrac{1}{2}$. None of the algorithms, however, reaches the lower bound given by the mutual information between past and future (dotted red curve), which is given by $1+p\log_2 p + (1-p)\log_2 (1-p)$ (see Refs.~\cite{crutchfield1997statistical, gu2012quantum}). To reach this bound phases may be required~\cite{liu2018optimal}.}\label{f:memory}
\end{figure}

\noindent{\em 1. Symmetrically perturbed coin process: }
Figures~\ref{f:qis_fig1}a and \ref{f:qis_fig2}a show an example recently investigated in Ref.~\cite{gu2012quantum} and experimentally demonstrated in Refs.~\cite{ghafari2019interfering,palsson2017experimentally}. This simple two-state Markov chain can be interpreted as the $\epsilon$-machine of a coin in a box undergoing regular perturbations that induce the coin to flip with probability $p$ at each time step~\cite{gu2012quantum}. 
In this case we have $f(i,x) = x\in\{0,1\}$ and the transition probabilities are given by 
\BE
P(1|0) = P(0|1) &=& p,\label{e:p}\\
P(0|0) = P(1|1) &=& 1-p \label{e:1-p}.
\EE
The best classical algorithms known to-date need to save in memory one bit, encoding whether the previous state was 0 or 1 (see Fig.~\ref{f:memory}).  

Figure~\ref{f:qis_fig2}a shows a possible implementation of a quantum $\epsilon$-machine that can reduce the memory requirements (see Fig.~\ref{f:memory}) by relying on quantum causal states~\cite{ghafari2019interfering,palsson2017experimentally,gu2012quantum} (cf. Eq.~\eqref{e:amplitude})
\BE
\left|\xi_0\ket &=& \sqrt{1-p}\left|{0}\ket + \sqrt{p}\left|{1}\ket,\label{e:S1} \\ 
\left|\xi_1\ket &=& \sqrt{p}\left|{0}\ket + \sqrt{1-p}\left|{1}\ket. \label{e:S2}
\EE
These quantum causal states, Eqs.~\eqref{e:S1} and \eqref{e:S2}, can be prepared from the state $\left| 0\ket$ via a unitary
\be\label{e:Ux}
U_x = \begin{pmatrix} \sqrt{1-x} & \# \\ \sqrt{x} & \#\end{pmatrix}, 
\ee
with $x=p$ and $x=1-p$, respectively. Here the undetermined entries are irrelevant for the protocol; they are chosen so that the operations are unitary. 

The quantum protocol starts with the current causal state $\left|\xi_j\ket$, with $j\in\{0,1\}$, adds an ancilla qubit in state $\left|\xi_0\ket$, and then applies to the combined system a gate (see Fig.~\ref{f:qis_fig2}a)
\be\label{e:CNOT2}
\textrm{CNOT}^{(1,2)} = \left|0\ket\bra 0\right|\otimes\id + \left|1\ket\bra 1\right|\otimes\mathbb{X} ,
\ee
where $\id$ is the identity matrix and
\be
\mathbb{X} = \begin{pmatrix} 0 & 1 \\ 1 & 0 \end{pmatrix}. \label{e:X}\\
\ee
This yields a new combined state
\be\label{e:chi}
\left|\chi_j\ket = \mathrm{CNOT}^{(1,2)}\left|\xi_j\ket \left|\xi_0\ket,
\ee
where
\BE
\left|\chi_0\ket &=& \sqrt{1-p}\left|{0}\ket\left|\xi_0\ket + \sqrt{p}\left|{1}\ket\left|\xi_1\ket, \label{e:CNOT|S1>|S1>}\\ 
\left|\chi_1\ket &=& \sqrt{p}\left|{0}\ket\left|\xi_0\ket + \sqrt{1-p}\left|{1}\ket\left|\xi_1\ket. \label{e:CNOT|S2>|S1>}
\EE
By measuring the first qubit in the computational basis we get the desired statistics and the corresponding transition of the second qubit to the correct quantum causal state, so the protocol can be applied iteratively~\cite{ghafari2019interfering,palsson2017experimentally}. 

As the stationary state of the symmetrically perturbed coin process (see Fig.~\ref{f:qis_fig1}a) is ${\pi=(\tfrac{1}{2},\tfrac{1}{2})}$, the density matrix associated to the quntum $\epsilon$-machine is (see Eq.~\eqref{e:rho1})
\be\label{e:rho_causal}
\rho = \frac{1}{2}\left|\xi_0\ket\bra\xi_0\right| + \frac{1}{2}\left|\xi_1\ket\bra\xi_1\right|= \lambda_+\left|+\ket\bra +\right| + \lambda_-\left|-\ket\bra -\right| .
\ee
Here $\lambda_+ =\lambda\equiv \tfrac{1}{2} + \sqrt{p(1-p)}$ and $\lambda_- = 1-\lambda$ are the largest and smallest eigenvalues, respectively, and 
\be
\left| +\ket = \frac{1}{\sqrt{2}}\begin{pmatrix} 1 \\ 1\end{pmatrix},\hspace{0.5cm} \left| -\ket = \frac{1}{\sqrt{2}}\begin{pmatrix} 1 \\ -1\end{pmatrix},
\ee
are the corresponding eigenvectors.

The quantum statistical memory is given by
\be\label{e:entropy_ex}
S_q(\lambda) = -\lambda\log_2\lambda - (1-\lambda)\log_2 (1-\lambda) \leq \log_2 2 = 1,
\ee
which is strictly smaller than the memory required for the classical $\epsilon$-machine (see Fig.~\ref{f:memory}), except when $p=0$ and $p=1$ where both $\epsilon$-machines require one bit, or $p=\tfrac{1}{2}$ where both $\epsilon$-machines require zero bits. So, if represented in terms of $\left| +\ket$ and $\left| -\ket$, the quantum $\epsilon$-machine can generate $M$ independent stochastic trajectories while keeping in memory only $M S_q(\lambda)$ qubits.

\

\noindent{\em 2. Post-processed perturbed coin process:}
Figures~\ref{f:qis_fig1}b and \ref{f:qis_fig2}b show an example, recently investigated in Ref.~\cite{thompson2018causal} and experimentally demonstrated in Ref.~\cite{ghafari2018single}, of a three-state Markov chain. This can be interpreted as the $\epsilon$-machine of a suitably post-processed perturbed coin process. See Refs.~\cite{thompson2018causal, ghafari2018single} for details on such an interpretation. Here we are only interested in how quantum protocols can generate samples from this Markov chain using a smaller topological memory than the best known classical algorithms. Indeed, there are quantum $\epsilon$-machines~\cite{thompson2018causal,ghafari2018single} that can simulate this Markov chain while keeping in memory just a single qubit, instead of a qutrit or two qubits as it may appear necessary for simulating a three-state system (see Fig.~\ref{f:qis_fig2}b). 

Following Sec.~\ref{s:quantum}, we can introduce an orthonormal basis $\{\left|0^\prime\ket ,\left|1^\prime\ket ,\left|2^\prime\ket\}$ (notice the primes) representing the deterministic causal states of the classical $\epsilon$-machine (see Eq.~\eqref{e:causal} and Fig.~\ref{f:qis_fig1}b). We can then define quantum causal states in terms of these as follows (see Eq.~\eqref{e:amplitude} and Fig.~\ref{f:qis_fig2}b): 
\BE
\left| \xi_0\ket &=& \sqrt{1-p}\left|0^\prime\ket + \sqrt{p}\left|2^\prime\ket =\left|0\ket,\label{e:|0'>}\\
\left| \xi_1\ket &=& \sqrt{q(1-p)}\left|0^\prime\ket + \sqrt{1-q}\left|1^\prime\ket + \sqrt{p q}\left|2^\prime\ket \label{e:|1'>}\\
&=& \sqrt{q}\left|0\ket + \sqrt{1-q} \left|1\ket ,\nonumber\\
\left| \xi_2\ket &=& \left|1^\prime\ket = \left|1\ket .\label{e:|2'>} 
\EE
Here we have also introduced a change of representation in terms of a new single qubit basis $\{\left|0\ket ,\left|1\ket \}$ (with no primes), where~\cite{thompson2018causal,ghafari2018single}
\BE
\left| 0\ket &=& \sqrt{1-p}\left|0^\prime\ket + \sqrt{p}\left|2^\prime\ket, \label{e:|0>} \\
\left|1\ket &=&\left|1^\prime\ket .\label{e:|1>}
\EE
Equations~\eqref{e:|0'>}-\eqref{e:|2'>} show the three quantum causal states $\left|\xi_i\ket$, for $i=0,1,2$, can indeed be written in terms of one single qubit, as stated above.

A quantum protocol~\cite{thompson2018causal} (see Fig.~\ref{f:qis_fig2}b) to sequentially generate samples from the post-processed perturbed coin process in Fig.~\ref{f:qis_fig1}b starts with the current causal state $\left|\xi_j\ket$, with $j\in\{0,1,2\}$, represented in terms of the qubit basis $\{\left|0\ket ,\left|1\ket \}$, 
along with two ancillary qubits, both in state $\left|0\ket$. The following unitary operations are then succesively applied to the three-qubit state $\left| \xi_i \ket\left| 0\ket\left| 0\ket$ (see Fig.~\ref{f:qis_fig2}b):
\BE
\neg\mathrm{CU}_p^{(1,3)} &=& \left|0\ket\bra 0\right| \otimes\id \otimes U_p + \left|1\ket\bra 1\right| \otimes\id\otimes\id,\label{e:-CUp} \\ 
\mathrm{CU}_{1-q}^{(1,2)} &=& \left|0\ket\bra 0\right| \otimes\id \otimes \id + \left|1\ket\bra 1\right| \otimes U_{1-q}\otimes\id, \label{e:CUq}\\
\mathrm{CNOT}^{(3,2)} &=& \id \otimes \id\otimes \left|0\ket\bra 0\right| + \id \otimes\mathbb{X}\otimes\left|1\ket\bra 1\right| , \label{e:CNOT}
\EE
where $U_p$ and $U_{1-q}$ are obtained by setting $x=p$ and $x=1-q$, respectively, in Eq.~\eqref{e:Ux}. We emphasize that the unspecified entries $\#$ in Eq.~\eqref{e:Ux} are not relevant for this protocol and are chosen such that $U_p$ and $U_{1-q}$ are unitary. 

Using the full unitary, 
\be
U = \mathrm{CNOT}^{(3,2)}\mathrm{CU}_{1-q}^{(1,2)}\neg\mathrm{CU}_p^{(1,3)},
\ee
built from these operators we obtain
\BE
U\left|\xi_0\ket\left|0\ket\left|0\ket &=& \left|0\ket (\sqrt{1-p}\left|0\ket\left|0\ket +\sqrt{p}\left|1\ket\left|1\ket ),\label{e:U0}\\
U\left|\xi_1\ket\left|0\ket\left|0\ket &=& \sqrt{q}\left|0\ket (\sqrt{1-p}\left|0\ket\left|0\ket +\sqrt{p}\left|1\ket\left|1\ket ) \\
& & + \sqrt{1-q}\left|1\ket (\sqrt{q}\left|0\ket +\sqrt{1-q}\left|1\ket )\left|0\ket ,\nonumber\label{e:U1}\\
U\left|\xi_2\ket\left|0\ket\left|0\ket &=& \left|1\ket (\sqrt{q}\left|0\ket +\sqrt{1-q}\left|1\ket )\left|0\ket .\label{e:U2}
\EE
A measurement of the first and third qubits in the qubit basis $\{\left|0\ket ,\left|1\ket\}$ (see Eqs.~\eqref{e:|0>} and \eqref{e:|1>}) is then performed. Let  $y_1\in\{0,1\}$ and $y_3\in\{0,1\}$ denote the corresponding values obtained. The probability of observing both $y_1=1$ and $y_3=1$ is zero. The combination $x=y_1+ 2 y_3\in\{0,1,2\}$ yields the three possible outputs of the post-processed perturbed coin process (see Fig.~\ref{f:qis_fig1}b) with the correct probabilities. The second qubit transitions to the correct quantum causal state (see Eqs.~\eqref{e:|0'>}-\eqref{e:|2'>}), so the protocol can be applied iteratively.

\section{Results}\label{s:results}
\subsection{Quantum-inspired memory-enhanced sampling}
\subsubsection{General considerations}
The quantum protocols described in Sec.~\ref{s:quantum} were based on the amplitude encoding of the transition probabilities of the corresponding Markov chains, Eq.~\eqref{e:amplitude} (see Fig.~\ref{f:qis_fig1}). However, Refs.~\cite{realpe2017modeling, realpe2018cognitive} show that the quantum dynamics of such phaseless quantum states can be mathematically reproduced via classical belief propagation, at least in the examples studied here (see Sec.~\ref{s:sqrt_bp} and Appendix~\ref{s:bp})---this only holds at the time scale set by the gates, not for any arbitrarily short time scale. It is then natural to ask whether such quantum memory-enhanced algorithms could be implemented classically. While belief propagation seems to be a strong candidate, there are some caveats that we expect can be resolved in the near future, as we will discuss in Appendix~\ref{s:bp}.

We will therefore introduce a different approach here. We will show that by exploiting some classical features implicit in the quantum protocols described in Sec.~\ref{s:quantum}, we can design classical algorithms more memory-efficient than the best classical algorithms known to date, i.e., $\epsilon$-machines. 

For simplicity, we will first discuss in Sec.~\ref{s:c-examples} the two examples we have been dealing with (see Figs.~\ref{f:qis_fig1} and \ref{f:qis_fig2}) and afterwards, in Sec.~\ref{s:extensions}, we discuss how to develop general quantum-inspired memory-enhanced stochastic algorithms. In Sec.~\ref{s:c-examples}, we will first introduce a classical algorithm for the symmetrically perturbed coin process (see Fig.~\ref{f:qis_fig1}a) that requires much less statistical memory than the corresponding $\epsilon$-machine, though it does not reach the same memory gains of the corresponding quantum protocol. 
Interestingly, as in the quantum protocol, the memory requirements of this classical memory-enhanced algorithm vary continuously along all the range of values of $p$, avoiding the discontinuous jump of the $\epsilon$-machine at $p=\tfrac{1}{2}$ (see Fig.~\ref{f:memory})---the slope of the curve, however, jumps discontinuously at $p=\tfrac{1}{2}$ while that of the quantum $\epsilon$-machine varies continuously. To achieve this, analogous to the quantum $\epsilon$-machine, the algorithm operates at the stationary state and at the ensemble level, i.e., on a large number of parallel independent samples. Furthermore, the algorithm exploits the eigendecomposition of the matrix of transition probabilities and an operational interpretation, introduced in Refs.~\cite{realpe2017modeling,realpe2018cognitive}, of negative numbers arising in a suitable decomposition of the vector of probabilities.

Next, we will introduce a classical algorithm for the post-processed perturbed coin process (see Fig.~\ref{f:qis_fig1}b) that has the same memory gains as the corresponding quantum protocol in Fig.~\ref{f:qis_fig2}b. More precisely, this algorithm needs to keep in memory only a single bit, instead of a trit as the corresponding $\epsilon$-machine (see Fig.~\ref{f:qis_fig1}b). To achieve this, we introduce {\em stochastic} causal states, i.e., classical mixtures of the `deterministic' causal states of the corresponding $\epsilon$-machine. Such mixtures do a similar job as the quantum superpositions of the amplitude-encoded quantum $\epsilon$-machines. We may refer to such classical protocols as stochastic $\epsilon$-machines.

In Sec.~\ref{s:extensions}, we will introduce general quantum-inspired memory-enhanced algorithms which, however, do not exploit square roots of probabilities. In Sec.~\ref{s:sqrt_bp} and Appendix~\ref{s:bp} we discuss how belief propagation can naturally lead to classical protocols formally similar to the corresponding quantum protocols. In particular, square roots of probabilities naturally arise in such protocols. However, we point out some caveats that we hope can be resolved in the near future, which would lead to a full algorithmic interpretation of square roots of probabilities with its corresponding additional memory gains. 


\subsection{Examples}\label{s:c-examples} 
\begin{figure}
\centering
\includegraphics[width=\columnwidth]{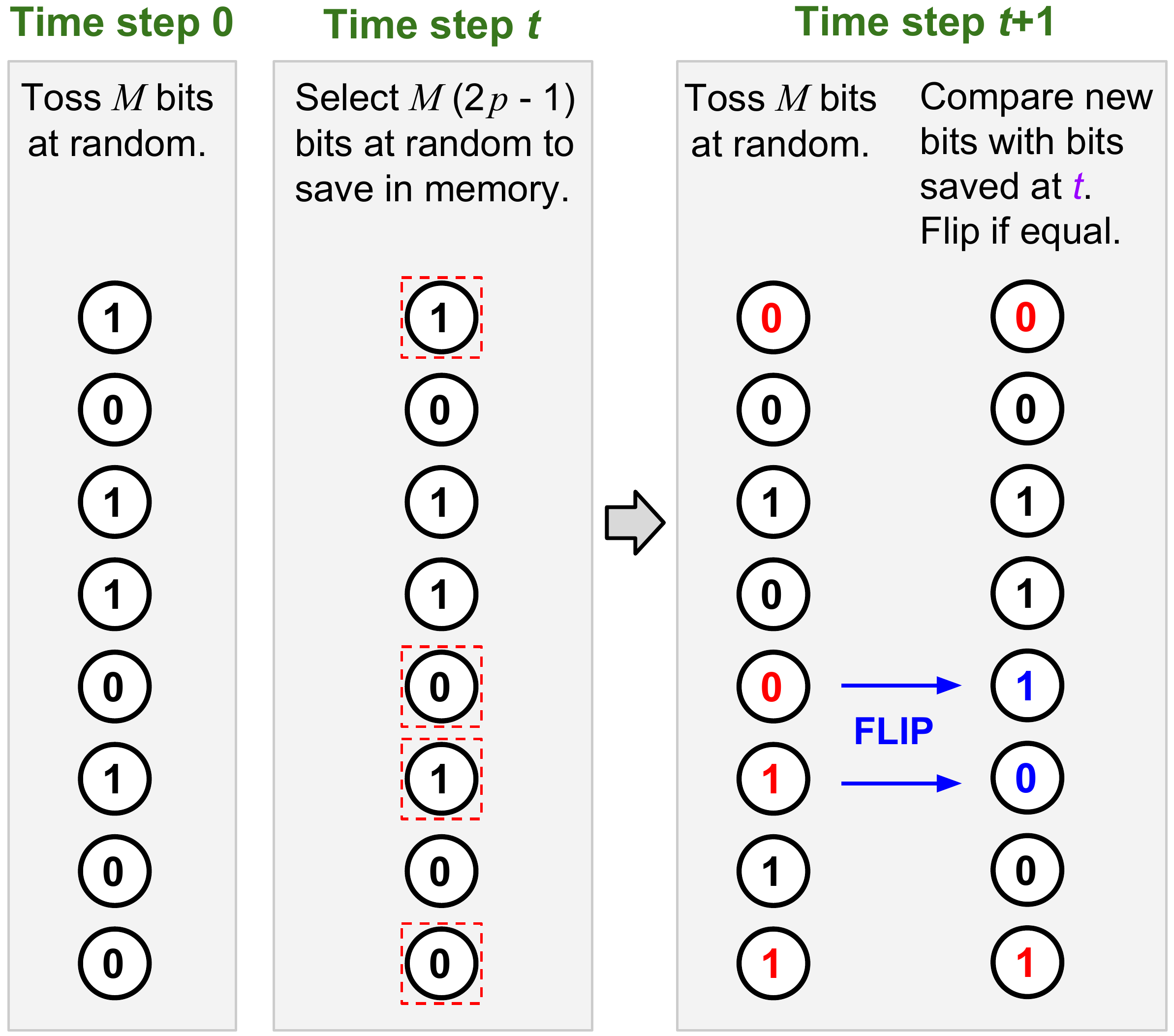}
\caption{{\em {New memory-enhanced classical algorithm:}} The reduction in statistical memory of quantum protocols like that in Fig.~\ref{f:qis_fig2}a takes place at the stationary state and manifests only at the ensamble level. Here we illustrate a classical algorithm for the symmetrically perturbed coin process that also operates at the stationary state and substantially reduces memory requirements at the ensamble level---here we assume $p>\tfrac{1}{2}$. The stationary state in this case is $\pi=(\tfrac{1}{2}, \tfrac{1}{2})$. We first prepare the stationary state by tossing $M\gg 1$ coins at random (left, time step 0). We then select $M |2p-1|$ of those coins and save their state in memory (center, generic time step $t$). To generate the next $M$ samples we can prepare again the stationary state by tossing again $M$ coins at random (right, time step $t+1$), which requires no memory about the past. We then scan through the $M |2p-1|$ coins whose state was saved in memory at time step $t$ and flip those that are equal, in case $p\geq\tfrac{1}{2}$, or flip those that differ, in case $p\leq\tfrac{1}{2}$. Since this will flip about half of the $M |2p-1|$ scanned coins, both from 0 to 1 and from 1 to 0, this conserves the stationary distirbution $\pi = (\tfrac{1}{2},\tfrac{1}{2})$. Moreover, this generates just enough bias to obtain the correct sufficient statistics characterizing a Markov chain, i.e., the stationary state and the frequencies characterizing all transition probabilities.}\label{f:qis_fig4}
\end{figure}

\subsubsection{Symmetrically perturbed coin process}\label{s:symm}
Here we introduce a classical memory-enhanced algorithm to sample from the symmetrically perturbed coin process that requires less statistical memory than the corresponding $\epsilon$-machine~\cite{gu2012quantum}
---though the algorithm does not reach the performance of the  quantum protocol.  
The lower statistical memory required by the quatum $\epsilon$-machine 
is based on the eigen-representation of the corresponding density matrix $\rho$. 
The eigen-representation of the density matrix $\rho$ associated to the symmetrically perturbed coin process is given by the right-most side of Eq.~\eqref{e:rho_causal}. When the process is completely random, i.e., $p=\tfrac{1}{2}$, we have $\lambda = 1$ and $\rho = \left| +\ket\bra +\right|$. In this case, the first term in the eigendecomposition of $\rho$ is associated to completely random, memoryless behaviour. So, when $p\neq\tfrac{1}{2}$, the second term in the eigendecomposition of $\rho$, which has negative elements, should somehow reintroduce the Markovian memory. 

The situation is similar to the one described in Refs.~\cite{realpe2017modeling,realpe2018cognitive} to operationally interpret negative numbers in a decomposition of general probability vectors, in this case the vector $(1-p,p)$ associated to a simple coin-toss process. Following Refs.~\cite{realpe2017modeling,realpe2018cognitive}, this probability vector can be written in a way similar to the eigendecomposition of $\rho$ in Eq.~\eqref{e:rho_causal}, namely
\be\label{e:neg_prob}
\begin{pmatrix} 1-p \\ p \end{pmatrix} = \frac{1}{2}\begin{pmatrix} 1 \\ 1 \end{pmatrix} + \frac{1-2p }{2}\begin{pmatrix} 1 \\ -1 \end{pmatrix} .
\ee

As before, the first vector can be interpreted as tossing coins uniformly at random, while the negative numbers in the second vector can be interpreted as a sort of `correction' (see Refs.~\cite{realpe2017modeling,realpe2018cognitive} and Sec.~\ref{s:extensions} below for the general version of this idea).  
More explicitly, the decomposition of the probability vector in Eq.~\eqref{e:neg_prob} could be read algorithmically as follows: (i) With probability one, toss coin uniformly at random (first vector); (ii) If $p<\tfrac{1}{2}$ and coin is in state $\left|1\ket =(0,1)^T$, with probability $1-2p$ flip the coin; (iii) If $p>\tfrac{1}{2}$ and coin is in state $\left|0\ket =(1,0)^T$, with probability $2p-1$ flip the coin.  The flipping in parts (ii) or (iii) would bias the uniform distribution of coins obtained in (i) just enough to get the correct statistics of the coin toss. 

Figure~\ref{f:qis_fig4} describes an algorithm that applies this idea to the case of the symmetrically perturbed coin process. Such an algorithm requires to save in memory only $|2p-1|\leq 1$ bits per sample instead of the $1$ bit of the best classical algorithm known to date (see Fig.~\ref{f:qis_fig2}c). In this, we first toss $M\gg 1$ coins at random and save the state of $M|2p-1|\leq M$ of them in memory. This would yield a set of samples from the stationary distribution $\pi = (\tfrac{1}{2},\tfrac{1}{2})$. To generate the next set of $M$ samples we toss again $M$ coins at random, which require zero memory, and then compare the $M|2p-1|$ coins saved in memory. If $p > \tfrac{1}{2}$ ($p<\tfrac{1}{2}$) we flip those coins whose state equals (differs from) the state before. This `correction' produces just enough bias to recover the correct statistics of the Markov chain, i.e., its transition probabilities, while respecting the stationary state $\pi = (\tfrac{1}{2},\tfrac{1}{2})$.

\subsubsection{Post-processed perturbed coin process}\label{s:post}
Here we introduce a memory-enhanced classical algorithm to sample from the post-processed perturbed coin process (see Fig.~\ref{f:qis_fig1}b) that requires the same topological memory of the corresponding quantum $\epsilon$-machine (see Fig.~\ref{f:qis_fig2}b), i.e., it requires to keep in memory only a bit instead of the trit required by the classical $\epsilon$-machine. The idea is to encode information stochastically by defining (classical) {\em stochastic} causal states in analogy with Eqs.~\eqref{e:|0'>}-\eqref{e:|2'>} as
\BE
\left| C_0\ket &=& \left|0\ket,\label{e:C0}\\
\left| C_1\ket &=& {q}\left|0\ket + (1-q) \left|1\ket ,\\
\left| C_2\ket &=& \left|1\ket.\label{e:C2}
\EE
As before, here the ket notation just refers to standard real vector notation. The stochastic causal state $\left|C_1\ket$ is uncertain; we do not have full knowledge about it. Yet we will see that this would allow us to obtain the very same memory savings as the corresponding quantum $\epsilon$-machine in Fig.~\ref{f:qis_fig2}b. In a sense, there seems to be computational value in knowing less.

Similarly, we can write the (classical) stochastic analogs of Eqs.~\eqref{e:U0}-\eqref{e:U2} as
\BE
\left|C_0\ket\left|0\ket\left|0\ket \rightarrow &\left| 0\ket [(1-p)\left|0\ket\left|0\ket +{p}\left|1\ket\left|1\ket ],\label{e:F0}\\
\left|C_1\ket\left|0\ket\left|0\ket \rightarrow &q\left| 0\ket [(1-p)\left|0\ket\left|0\ket +{p}\left|1\ket\left|1\ket ] + \\
&+ (1-q)\left|1\ket [q\left|0\ket +(1-q)\left|1\ket ]\left|0\ket ,\nonumber\\
\left|C_2\ket\left|0\ket\left|0\ket \rightarrow &\left| 1\ket (q\left|0\ket +(1-q)\left|1\ket )\left|0\ket .\label{e:F2}
\EE
Indeed, let  $y_1\in\{0,1\}$ and $y_3\in\{0,1\}$ denote the values obtained after observing the first and third stochastic bits, respectively. The probability of observing both $y_1=1$ and $y_3=1$ is zero. The combination $x=y_1+ 2 y_3\in\{0,1,2\}$ yields the three possible outputs of the post-processed perturbed coin process (see Fig.~\ref{f:qis_fig1}b) with the correct probabilities. The second stochastic bit transitions to the correct stochastic causal state (see Eqs.~\eqref{e:C0}-\eqref{e:C2}), so the protocol can be applied iteratively. Since everything is real and non-negative, we can readily build a classical stochastic algorithm that implements the transitions in Eqs.~\eqref{e:F0}-\eqref{e:F2}. 

For instance, if the initial output state is $x = 0$ or $x=2$, we set the first stochastic bit to $s^{(1)}=0$ or $s^{(1)}=1$, respectively. Otherwise, we set it to $s^{(1)}=0$ or $s^{(1)}=1$ with probabilities $q$ and $1-q$, respectively. To generate the next output, we do the following iteration: if $s^{(1)}=0$ we set both the second and third stochastic bits to zero or one, i.e., $s^{(2)}=0=s^{(3)}$ or $s^{(2)}=1=s^{(3)}$, with probabilities $1-p$ and $p$, respectively. Otherwise, if $s^{(1)}=1$ we set $s^{(1)}=0$ or $s^{(1)}=1$ with probabilities $q$ and $1-q$, respectively; furthermore, we set $s^{(3)}=0$. We then output $x=s^{(1)}+2\, s^{(3)}$ and update the memory stochastic bit as $s^{(1)} = s^{(2)}$. By iterating this procedure we get a sequence of outputs $x$ that is a sample trajectory of the post-processed perturbed coin process (see Fig.~\ref{f:qis_fig1}a). As in the case of the quantum $\epsilon$-machine, here we only need to keep in memory the first stochastic bit, $s^{(1)}$, which at the end of each iteration will always be in one of the stochastic causal states, Eqs.~\eqref{e:C0}-\eqref{e:C2}.

\subsection{Extensions}\label{s:extensions}
\subsubsection{General considerations}\label{s:gral}

Here we highlight some of the general concepts underlying the examples discussed in Sec.~\ref{s:c-examples}. We first introduce a general quantum-inspired stochastic algorithm that extends the one introduced in Fig.~\ref{f:qis_fig4} to general Markov chains. As the one introduced in Fig.~\ref{f:qis_fig4}, this algorithm also exploits the interpretation of negative numbers introduced in Refs.~\cite{realpe2017modeling,realpe2018cognitive}. Afterwards, we discuss the concept of general stochastic causal states that extends the example of the post-processed perturbed coin process to general Markov chains and $\epsilon$-machines. None of these generaliations exploits square roots of probabilities. This possibility is discussed briefly in Sec~\ref{s:sqrt_bp} and in more detail in Appendix~\ref{s:bp}.

To begin, the algorithm described in Sec.~\ref{s:symm} (see Fig.~\ref{f:qis_fig4}) can be interpreted as an eigendecomposition of the matrix of transition probabilities $T_{j\, j^\prime} = P(j^\prime |j)$. According to the Perron-Frobenius theorem for irreducible stochastic matrices, the stationary state corresponds to the (left)-eigenvector $\bra\pi\right|$ with eigenvalue equal to one, i.e., $\bra\pi\right| T= \bra\pi\right|$, which is the eigenvalue with largest absolute value. In this sense, sampling from the stationary state, which is the first step in the algorithm described in Fig.~\ref{f:qis_fig4}, yields the main contribution to generate stochastic trajectories with the correct statistics. The remaining eigenvectors yield `corrections', which may involve negative numbers as in the example described in Fig.~\ref{f:qis_fig4}, whose relevance is given by the magintude of their associated eigenvalues. 

In principle, these ideas can be further generalized by using eigendecompositions of more general matrices of transition probabilities and interpreting the negative numbers that may arise in a similar way as we have done in the example described in Fig.~\ref{f:qis_fig4}. Such potential generalizations should take into account that for non-symmetric matrices, unlike the example described in Fig.~\ref{f:qis_fig4}, the right and left eigenvectors may be different. In Sec.~\ref{s:gqi} we discuss a general quantum-inspired memory-enhanced algorithm based on this idea. In Sec.~\ref{s:stoch_causal} we discuss the notion of general stochastic causal states that generalizes the ideas exploited in the example studied in Sec.~\ref{s:post}.  Finally, in Sec.~\ref{s:sqrt_bp} and Appendix~\ref{s:bp} we discuss how square roots might also be exploited via belief propagation protocols.

\subsubsection{General quantum-inspired algorithm with enhanced statistical memory}\label{s:gqi}

Here we describe how taking into account only one left- and right-eigenvector, i.e., those associated to the stationary state, already leads to a general quantum-inspired memory-enhanced stochastic algorithm. Consider an $(N+1)$-state Markov chain with state space $\mcS$ specified by the matrix of transition probabilities (see Appendix~\ref{s:exqi} and Fig.~\ref{f:general} for a detailed example)
\be\label{e:T}
T = \left|\id\ket\bra\pi\right| + \Delta ,
\ee 
where $\bra\pi\right| = (\pi_0,\dotsc ,\pi_N)$ is the stationary state, $\left|\id\ket$ is the corresponding right-eigenvector, i.e., the $(N+1)$-dimensional vector with all entries equal to one, and $\Delta = T-\left|\id\ket\bra\pi\right|$ is composed of all the eigenvalues and eigenvectors different from $\bra\pi\right|$ and $\left|\id\ket$. For simplicity, we are asuming ergodicity so there is a unique stationary distribution.

For each $j\in\mcS$, let 
\BE
\mcS_j^- &=& \left\{ i\in\mcS : \bra j|\Delta| i\ket < 0 \right\},\\
\mcS_j^+ &=& \left\{ i\in\mcS : \bra j|\Delta| i\ket > 0 \right\}. 
\EE
It is useful to define the positive fractions
\be\label{e:f_j}
f_j = \max\left[-\frac{\bra j|\Delta| i\ket}{\pi_{i}} : i\in\mcS_j^- \right],
\ee
and the positive ratios
\BE
r^-_{j:i\to} &=& -\frac{\bra j|\Delta |i\ket}{f_j\pi_i},\hspace{0.5cm}\textrm{  for  } i\in\mcS_j^- , \label{e:r-} \\ 
r^+_{j:\to i^\prime} &=& \frac{1}{Z_j}{\bra j|\Delta |i^\prime\ket},\hspace{0.5cm}\textrm{  for  } i^\prime\in\mcS_j^+ , \label{e:r+}
\EE
where $Z_j$ is a normalization constant enforcing $\sum_{i^\prime\in\mcS_j^+} r_{j:\to i^\prime}^+ = 1$. 

To see that $f_j\leq 1$ we can show that all $-\bra j|\Delta| i\ket / {\pi_{i}} \leq 1$. Indeed, assume by contradiction that $-\bra j|\Delta| i\ket / {\pi_{i}} > 1$ so, from Eq.~\eqref{e:T}, we get $-\bra j| T |i\ket + \pi_i > \pi_i$, i.e., $-\bra j| T |i\ket > 0$. This is a contradiction since $\bra j \right|T$ is a probability distribution. Similarly, since by definition $f_j$ is the maximum of all terms $-\bra j|\Delta| i\ket / {\pi_{i}}$, then all ratios $r^-_{j:i\to}\leq 1$. Finally, all ratios $r_{j:\to i^\prime}^+\leq 1$ due to the normalization constant $Z_j$. So, all $f_j$, $r^-_{j:i\to}$, and $r_{j:\to i^\prime}^+$ can be interpreted as probabilities.

A general quantum-inspired memory-enhanced stochastic algorithm can be designed as follows. Let $s_\ell^t\in\mcS$, with $\ell = 1,\dotsc , M$, denote the $\ell$-th sample generated at time step $t$. For easy of reference, we will keep this ordering of samples throughout. If at time step $t$ the algorithm produces output $j$ we save it in memory with probability $f_j$. At the end of time step $t$ we have generated $M$ outputs, $\{s_\ell^t\}_{\ell =1}^M$, distributed according to the stationary state $\bra\pi\right|$, and have saved in memory on average $m_j=f_j\pi_j M$ samples in state $\bra j\right|$, with $j\in\mcS$ and $\pi_j = \bra\pi |j\ket$. Let
\be
\mathcal{M}_j^t = \left\{\ell : s_\ell^t = j \textrm{ was saved in memory}  \right\},
\ee
denote the set of indexes of the samples $s_\ell^t$ with value $j\in\mcS$ that are saved in memory at time $t$. After this process we have a total of $m = \sum_{j\in\mcS}m_j$ samples saved. 

To generate the $\ell$-th sample, $s_\ell^{t+1}$, at time step $t+1$ we first generate a sample from the stationary state $\bra\pi\right|$ as an intermediate stage. If the $\ell$-th sample was not saved at time step $t$, i.e. $\ell\notin\bigcup_{j\in\mcS}\mcM_j^t$, we just output the new sample $s_\ell^{t+1}$. Now, suppose the $\ell$-th sample, $s_\ell^t=j$, was indeed saved at time step $t$ with value $j\in\mcS$, i.e., $\ell\in\mcM_j^{t}$. Suppose also that the new $\ell$-th sample, $s_\ell^{t+1} = i$, generated at the intermediate stage has value $i$. If $i\in\mcS_j^+$ we just output the sample. Otherwise, if $i\in\mcS_j^-$ we change it into $i^\prime\in\mcS_j^+$ with probability 
\be\label{e:r->}
r_{j:i\to i^\prime} = r_{j:i\to }^- r_{j:\to i^\prime}^+ .
\ee
This process is repeated $M$ times.

To see that this algorithm indeed generates samples with the right statistics, consider the probability $\mathcal{P}_{j\to i}$ to obtain a sample $s_\ell^{t+1} = i$ at time $t+1$ corresponding to a sample $s_\ell^t = j$ at time step $t$. In the intermediate stage we obtain  $s_\ell^{t+1} = i$ with probability $\pi_i$. The probability that both $s_\ell^t=j$ was saved in memory at time step $t$ and $s_\ell^{t+1}=i$ at the intermediate stage is $\pi_i f_j$. Now, if $s_\ell^t=j$ was saved in memory and $s_\ell^{t+1} = i\in\mathcal{S}^-_j$ at the intermediate stage, the value of the sample $s_\ell^{t+1}$ will change into any $i^\prime\in\mcS_j^+$ with probability $\sum_{i^\prime\in\mcS_j^+} r_{j:i\to i^\prime}$. This leads to the relation
\be
\mathcal{P}_{j\to i} = \pi_i - \pi_i f_j \sum_{i^\prime\in\mcS_j^+} r_{j:i\to i^\prime},
\ee
which using Eqs.~\eqref{e:r-}-\eqref{e:r->} yields $\mathcal{P}_{j\to i} = \pi_i + \bra j|\Delta| i\ket$, or $\mathcal{P}_{j\to i} = \bra j|T| i\ket$, as expected. 

If $i\in\mcS_j^+$ instead, the sample $s_\ell^{t+1} = i$ does not change. However, we have to take into account all possible transitions into $i$ from all other samples $s_{\ell^\prime}^{t+1}$ that at the intermediate stage satisfy $s_{\ell^\prime}^{t+1} = i^\prime$, with $\ell^\prime\in\mcM_j^t$, $\ell^\prime\neq\ell$, and $i^\prime\in\mcS^-_j$. Since at the intermediate stage $s_{\ell^\prime}^{t+1} = i^\prime $  with probability $\pi_{i^\prime}$ and $s_{\ell^\prime}^t = j$ was saved in memory with probability $f_j$, both events happen with probability $f_j\pi_{i^\prime}$. Now, the total probability that any of the samples $s_{\ell^\prime}^{t+1} = i^\prime$, with $i^\prime\in\mcS_j^-$, corresponds to a sample $s_\ell^t = j$ saved in memory, i.e., $\ell^\prime\in\mcM_j^t$, and transitions into $i$ is $\sum_{i^\prime\in\mcS_j^-}\pi_{i^\prime}f_j r_{j:i^\prime\to i}$. This leads to the relation
\be
\mathcal{P}_{j\to i} = \pi_i + \sum_{i^\prime\in\mcS_j^-}\pi_{i^\prime} f_j  r_{j:i^\prime\to i},
\ee
which using Eqs.~\eqref{e:r-}-\eqref{e:r->} yields
\be
\mathcal{P}_{j\to i} = \pi_i - \frac{\bra j|\Delta | i\ket}{Z_j} \sum_{i^\prime\in\mcS_j^-} \bra j|\Delta | i^\prime\ket . 
\ee
To change the sum above into a sum over elements of $S_j^+$ we can use the identity
\be
\begin{split}
\sum_{i^\prime\in \mcS_j^+}\bra j| \Delta|i^\prime\ket  + \sum_{i^\prime\in \mcS_j^-}\bra j| \Delta|i^\prime\ket & = 
\sum_{i^\prime\in \mcS_j^+\cup\mcS_j^-}\bra j| \Delta|i^\prime\ket \\
&= \sum_{i^\prime\in \mcS}\left[\bra j |T|i^\prime\ket -\pi_{i^\prime}\right] = 0.
\end{split}
\ee
Here we have extended the sum from $\mcS_j^+ \cup\mcS_j^-$ to $\mcS$ since we can add all $\bra j|\Delta| i^\prime\ket = 0$ without changing the sum. We have also taken into account that both $\pi_i$ and $\bra j|T|i\ket$ are probability distributions over $i$. Using this identity and Eq.~\eqref{e:r+} we finally obtain $\mathcal{P}_{j\to i} = \pi_i + \bra j|\Delta|i\ket$, or $\mathcal{P}_{j\to i} = \bra j|T| i\ket$, as expected.

Again, we have built this in analogy with the quantum protocols that reduce statistical complexity (see Sec.~\ref{s:quantum}), which exploit the eigendecomposition of the density matrix to find an optimal quantum encoding of the $M$ samples. However, we have exploited here only the eigenvector corresponding to the stationary state. It may be possible to exploit all eigenvalues to build algorithms that require less memory. Nevertheless, like in the case of the quantum protocols, the price to pay for this would be the need to find and deal with the full eigensystem of the matrix of transition probabilities.

\subsubsection{General stochastic causal states}\label{s:stoch_causal}

Now, the concept of stochastic causal state introduced in Sec.~\ref{s:post} (see Eqs.~\eqref{e:C0}-\eqref{e:C2}) can in principle be generalized to systems of any dimension. For instance, in the case of a Markov chain specified by matrix of transition probabilities $T$, in analogy with Eqs.~\eqref{e:C0}-\eqref{e:C2},  we could define general stochastic causal states as
\be
\left| C_i \ket = \sum_{j\in\mcS}P(j|i)\left| j\ket . 
\ee
Or in the case of general $\epsilon$-machines, in analogy with Eq.~\eqref{e:amplitude}, we could define them as
\be
\left|C_i\ket = \sum_{x\in\mcA , j\in\mcS}P(x,j|i) \left|j\ket\left| x\ket ,
\ee
where $\{\left|j\ket\}_{j\in\mcS}$ is an orthonormal basis that represents the deterministic causal states $j\in\mcS$ of the classical $\epsilon$-machine, and $\{\left|x\ket\}_{x\in\mcA}$ is an orthonormal basis that represents the corresponding outputs $x\in\mcA$. 

We can in principle find a classical algorithm that requires a state space with the same dimensionality, $D_{\rm stoch}$, of the space spanned by the stochastic causal states, $\{\left| C_i\ket\}_{i\in\mcS}$. If $D_{\rm stoch}$ is smaller than the dimensionality $|\mcS|$ of the original state space, $\mcS$, we would obtain a quantum-inspired classical algorithm that requires less topological memory than the best classical counterpart known to date, i.e., the $\epsilon$-machine. This was the case of the post-processed perturbed coin process studied in Sec.~\ref{s:post}. This particular instance has the advantage that it does not involve negative numbers. This feature facilitates the representation in terms of classical stochastic causal states, Eqs.~\eqref{e:C0}-\eqref{e:C2}, and the corresponding transitions, Eqs.~\eqref{e:F0}-\eqref{e:F2}. In more general situations there may be negative numbers involved, which could in principle be interpreted operationally as we have done with the symmetrically-perturbed coin process (see Fig.~\ref{f:qis_fig4}) and its generalization in Sec.~\ref{s:gqi} (see also Appendix~\ref{s:exqi}).

\subsubsection{Square roots of probabilities and belief propagation}\label{s:sqrt_bp}

In this work we have discussed memory-enhanced quantum-inspired stochastic algorithms that, unlike the quantum protocols discussed in Sec.~\ref{s:quantum}, do not make use of square roots of probabilities. In some instances this has led to lower memory savings than those that can be achieved with the corresponding quantum protocols (see, e.g., Sec.~\ref{s:symm}). So, dealing with square roots of probabilities has the potential of achieving the same performance of quantum protocols in this cases. 

Indeed, what actually motivated us to search for such quantum-inspired algorithms was the observation that the classical algorithm known as belief propagation, when run on cycle- or chain-like classical graphical models, follows a dynamics which is mathematically similar to quantum dynamics when there are no phases involved~\cite{realpe2017modeling, realpe2018cognitive}. Interestingly, square roots of probabilities and the analogs of the Born rule arise naturally in this potentially more general framework. 

In particular, it is possible to build graphical models whose belief propagation dynamics is mathematically analogous to the quantum protocols discussed in Sec.~\ref{s:qex}.  There are, however, some caveats that we hope can be resolved in the near future. With this in mind, we present in Appendix~\ref{s:bp} a general discussion of these ideas along with the two graphical models associated to the two quantum protocols discussed in Sec.~\ref{s:qex}. This approach might in principle lead to classical algorithms where the amplitude encodings associated to the corresponding quantum protocols could be understood as organizing samples in squares, instead of lines as in Fig.~\ref{f:qis_fig4}.  Any required linear algebra manipulation would then be applied to the sides of such squares rather than to the full squares themselves. A full algorithmic interpretation of square roots of probabilities is left for future work.

\section{Conclusions}\label{s:conclusions}

Memory is a key computational resource for performing high-performance simulations, e.g., of large-scale complex systems or quantum computers. In this work we have shown that some quantum protocols~\cite{thompson2018causal,gu2012quantum,ghafari2018single,ghafari2019interfering,palsson2017experimentally} recently introduced for stochastic simulation, which can provide extreme memory advantages over the best classical protocols known to date, can actually be either implemented or approximated classically. In the former case we can design classical algorithms with the same memory reductions of the quantum counterparts. In the latter we can design algorithms that require less memory than the best classical algorithms known to date, yet they do not reach the same memory reductions of the quantum counterparts, so there is potentially still room for further improvement. 

One of the concepts involved is the encoding of information about the past of a system on probabilistic mixtures, even though such information is already known and so, in a sense, `deterministic'. We showed how this can lead to a reduction in the dimension of the state space, or topological memory, of the variables that need to be kept in memory in order to sequentially generate a stochastic trajectory. In short, there is value in knowing less. This is the stochastic parallel of the corresponding quantum protocols which reduce memory requirements by saving already observed information about the past on quantum superpositions.

Another concept involved is the decomposition of matrices of transition probabilities into the dominant left and right eigenvectors, i.e., those associated to the stationary state, and the residual. We can then sample first from the stationary state, which requires zero memory about the past, and then correct the samples according to the residual to recover temporal correlations. This may involve negative numbers that may arise in the decomposition. We have used an operational interpretation of such negative numbers introduced in recent work~\cite{realpe2017modeling,realpe2018cognitive}, to design classical sampling algorithms that require much less memory than the best known to date, i.e., the classical $\epsilon$-machines.

Furthermore, following Refs.~\cite{realpe2017modeling,realpe2018cognitive}, we have discussed how the classical message-passing algorithm known as belief propagation can lead to a dynamics mathematically similar to amplitude-encoded quantum protocols, where the phase degree of freedom does not play any role. In particular, square roots of probabilities arise naturally in this framework. Except for a very recent quantum protocol~\cite{ghafari2018single}, all quantum memory-enhanced protocols for stochastic simulation investigated to date, like the belief propagation counterparts, do not make use of quantum phases. Such restricted amplitude-encoded quantum protocols, though, can lead to extreme memory reductions~\cite{aghamohammadi2018extreme}. A full algorithmic interpretation of such square roots of probabilities, however, is left for future work.

Being classical, such quantum-inspired algorithms could be implemented in state-of-the-art high-performance computers. This could potentially enhance the study of complex systems and further raise the bar for near-term quantum computers to demonstrate `quantum supremacy'. 

To conclude, the possibility to classically implement protocols that were previously considered quantum raises the question: what is quantum? An intriguing possibility which arose in Refs.~\cite{realpe2017modeling, realpe2018cognitive} (see also Ref.~\cite{realpe2019can}), which motivated this work, is that the peculiar {\em classical} interaction of a physical agent, e.g., a robot or a scientist, with a classical experimental setup, when described from the perspective of the agent itself leads to a fully quantum-like dynamics. If this is so, it might be possible to extend the ideas introduced here to general quantum protocols that may include phase degrees of freedom, as those recently studied in Ref.~\cite{ghafari2018single}.


\section*{Contributions}

J.R.G. conceived the main ideas and derived all technical results. N.K. supervised the project and provided critical feedback to direct the research. J.R.G. wrote the manuscript in consultation with N.K.

\section*{Competing interests}

The authors declare no competing interests.



\clearpage

\newpage

\section*{Supplementary information}

\appendix

\section{Illustrative example of algorithm in Sec.~\ref{s:gqi}}\label{s:exqi}
\begin{figure}
\includegraphics[width = \columnwidth]{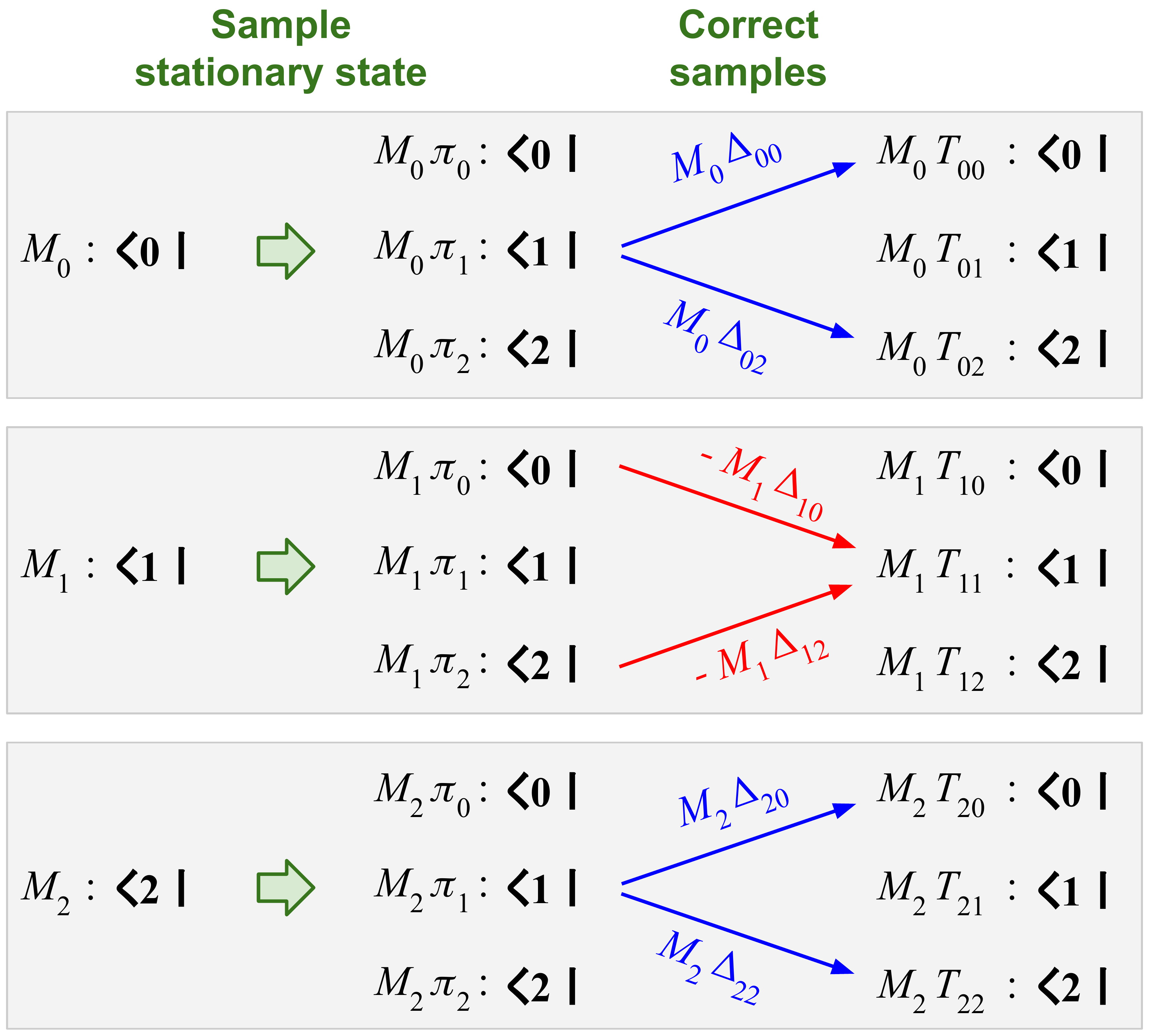}
\caption{{\em General memory-enhanced classical algorithm:} Here we illustrate the general version of the algorithm described in Fig.~\ref{f:qis_fig4}, using the Markov chain specified by transition matrix $T$ in Eq.~\eqref{e:T3}. Consider $M=18^2$ samples currently distributed according to the stationary state in Eq.~\eqref{e:pi3}, i.e., $M\bra\pi\right|\equiv (M_0, M_1, M_2) = 18\times (4, 9, 5)$ (first column). To generate the next samples, in an intermediate stage we first generate $M$ samples again from the stationary state, which requires zero memory (second column). This set of $M$ samples can be classified into three sets of $M_j$ samples, for $j\in\{0,1,2\}$, each distributed according to the stationary state, i.e., $M_j\bra\pi\right|$, which will be transformed into the new samples transitioning from current states $\bra j\right|$, respectively. We need $M_j T_{jk}$ samples, with $T_{jk}\equiv\bra j | T|k\ket$ and $k\in\{0,1,2\}$, transitioning from state $\bra j\right|$ to state $\bra k\right|$. In the intermediate stage we have $M_j\pi_k$, with $\pi_k = \bra\pi |k \ket$, instead. We can use vector $\Delta$ in Eq.~\eqref{e:D3} to `correct' the samples obtained in the intermediate stage (third column). Indeed, according to Eq.~\eqref{e:<j|T} we just need to add $M_j\Delta_{jk}$ to the $M_j\pi_k$ samples, with $\Delta_{jk}\equiv\bra j|\Delta |k\ket$. This addition can be considered as transitions from states $\bra k_-\right|$ with $\Delta_{jk_-}<0$ into states $\bra k_+\right|$ with $\Delta_{jk_+} >0$, where $k_+, k_-\in\{0,1,2\}$ (see Eqs.~\eqref{e:D03}-\eqref{e:D23}). Now, the number of samples that need be transformed can be much smaller than the total number of samples $M$. For instance, if $T_{jk} = \pi_k$ we would not need to transform any sample nor save anything in memory. In this example we need to save only $m_0^\ast\equiv\tfrac{1}{3}M\pi_0$, $m_1^\ast\equiv\tfrac{1}{2}M\pi_1$, and $m_2^\ast\equiv\tfrac{1}{3}M\pi_2$ samples  in states $\bra 0\right|$, $\bra 1\right|$, and $\bra 2\right|$, respectively, or a total of $\tfrac{5}{12} M < M$. If we do not want to differentiate between states, we can save $m^\ast\equiv 3\max\{m_0^\ast, m_1^\ast, m_2^\ast\} = \tfrac{3}{4}M$. Finally, instead of working with the total number of samples that need be `corrected' we can work with ratios, which can be interpreted as probabilities. This allows us to sequentially apply the `corrections', online, to each sample generated. So, contrary to what this figure may suggest, we only need to save $\tfrac{5}{12} M$,  or $\tfrac{3}{4}M$, samples in memory (see Sec.~\ref{s:extensions}). } \label{f:general}
\end{figure}

Here we discuss a specific three-state Markov chain to illustrate the general algorithm introduced in Sec.~\ref{s:gqi}. Instead of just replacing the specific values here into the general expressions introduced therein, we will rather work this example out from scratch to motivate the latter. Although this will necessarily lead to a much longer discussion, we hope this approach presents a complementary perspective that may help clarify any confusion that may have arisen in reading the general presentation in Sec.~\ref{s:gqi}.

Consider the three-state Markov chain specified by the transition probability matrix
\be\label{e:T3}
T = \begin{pmatrix} 
\frac{1}{3} & \frac{1}{3} & \frac{1}{3}\\
p & q & 1-p-q\\
\frac{1}{3} & \frac{1}{3} & \frac{1}{3}
\end{pmatrix} = \left|\id\ket\bra\pi\right| + \Delta .
\ee
Here we have written $T$ in terms of the left and right eigenvectors associated to the largest eigenvalue $\lambda = 1$, which are 
\be
\bra\pi\right| = \frac{1}{4-3 q}\begin{pmatrix} 1 + p -q, &  1, & 2(1-q) -p \end{pmatrix},\label{e:pi3pq}
\ee
and $\left|\id\ket = (1,1,1)^T$, respectively. The reverse relation of Eq.~\eqref{e:T3} is
\be\label{e:D3}
\Delta = T-\left|\id\ket\bra\pi\right|.
\ee 
The eigenvector $\bra\pi\right|$ in Eq.~\eqref{e:pi3pq} is the stationary state of the Markov chain. 

Although the Markov chain specified by the transition matrix $T$ in Eq.~\eqref{e:T3} could be described by just two causal states, it serves to illustrate the main ideas we want to discuss. We could use the same ideas for the corresponding two-level $\epsilon$-machine, but this would be a rather trivial example. This also serves to illustrate that we can improve not only on $\epsilon$-machines, but also on sub-obtimal algorithms that may arise in real-life applications, where finding the corresponding $\epsilon$-machine may be hard. Of course, this observation also applies for the orginial quantum protocols.

If the state of the system is $\bra j\right|$, with $j\in\{0,1,2\}$, its next state is drawn from the probability vector
\be\label{e:<j|T}
\bra j\right| T = \bra\pi\right| + \bra j\right|\Delta. 
\ee
In the spirit of Fig.~\ref{f:qis_fig4} we can generate samples by first sampling from the stationary state $\bra\pi\right|$, which is the same for all $j$, and afterwards correct the samples according to the second term in Eq.~\eqref{e:<j|T}, i.e., $\bra j\right|\Delta$.

For concreteness, let us consider the case when $p=\tfrac{1}{9}$ and $q=\tfrac{2}{3}$, which yields (see Eqs.~\eqref{e:pi3pq} and \eqref{e:D3})
\BE
\bra\pi\right| &=& \frac{1}{18}\begin{pmatrix} 4, & 9, & 5\end{pmatrix} =  \begin{pmatrix} \pi_0 ,\pi_1 ,\pi_2\end{pmatrix},\label{e:pi3}\\
\bra 0\right|\Delta &=& \frac{1}{18}\begin{pmatrix} 2, & -3, & 1\end{pmatrix}\label{e:D03},\\
\bra 1\right|\Delta &=& \frac{1}{18}\begin{pmatrix} -2, & 3, & -1\end{pmatrix}\label{e:D13},\\
\bra 2\right|\Delta &=& \frac{1}{18}\begin{pmatrix} 2, & -3, & 1\end{pmatrix},\label{e:D23}
\EE
and suppose we are generating $M=18^2$ samples in parallel (see Fig.~\ref{f:general}; cf. Fig.~\ref{f:qis_fig4}). The $M$ samples are at the stationary state, Eq.~\eqref{e:pi3}, so the number of samples in states $\bra j\right|$, with $j=0$, $j=1$ and $j=2$, is on average
\BE
M_0 &=& M\pi_0 = 18\times 4,\\
M_1 &=& M\pi_1 = 18\times 9,\\
M_2 &=& M\pi_2 = 18\times 5,
\EE
respectively. For easy of illustration, we will work with sample statistics as if they were population statistics, i.e., we will neglect fluctuations about the mean in the following discussion.

To generate the next set of samples, we first generate $M = 18^2$ new samples from the stationary state $\bra\pi\right|$, Eq.~\eqref{e:pi3}. This is equivalent to generate three subsets of samples of size $M_0$, $M_1$ and $M_2$, each distributed according to $\bra\pi\right|$ (see Fig.~\ref{f:general}). To recover the temporal correlations between the current and the next set of samples, we need to `correct' each subset of $M_j$ samples using vectors $\bra j\right|\Delta$, with $j=0,1,2$. Consider, for instance, the $M_0$ samples that transition from state $\bra j=0\right|$. Since these $M_0$ samples are distributed according to the stationary state, Eq.~\eqref{e:pi3}, on average $M_0\pi_0 = 4\times 4$ stay in state $\bra j=0\right|$,  $M_0\pi_1 = 4\times 9 $ transition to state  $\bra j=1\right|$, and $M_0\pi_2 = 4\times 5$ transition to state $\bra j=2\right|$. 

To `correct' these $M_0$ samples according to vector $\bra 0\right|\Delta$ in Eq.~\eqref{e:D03} we need to change some of the samples that are in states with negative entries in $\bra 0\right|\Delta$, i.e., state $\bra j=1\right|$, into states that have positive entries in $\bra 0\right|\Delta$ according to the ratios specified by these entries. So, for every three samples in state $\bra j=1\right|$ that change, two should change into state $\bra j=0\right|$ and one into state $\bra j=2\right|$. More precisely, since in this case the only negative entry in Eq.~\eqref{e:D03} is $\bra 0\right|\Delta\left|1\ket = -{3}/{18}$, we need to change on average $-M_0\bra 0\right|\Delta\left|1\ket = 4\times 3$ samples. On average $M_0\bra 0\right|\Delta\left|0\ket = 4\times 2$ should change into state $\bra j=0\right|$ and $M_0\bra 0\right|\Delta\left|2\ket = 4\times 1$ should change into state $\bra j=2\right|$. This recovers the correct statistics of the transitions from state $\bra j=0\right|$ for the simple reason that the corresponding vector of transition probabilities $\bra 0\right| T = \bra\pi\right| + \bra 0\right|\Delta$ is the sum of the stationary distribution vector and the `correction' vector (see Eq.~\eqref{e:<j|T} and general discussion in Sec.~\ref{s:gqi}). Although this `correction' generally take the $M_0$ samples out of the stationary state, all $M=M_0+M_1+M_2$ samples will globally remain at the stationary state once we also `correct' the samples that transition from states $\bra j=1\right|$ and $\bra j=2\right|$, as we will see below.

Now, the correction vector $\bra 1\right|\Delta$ in Eq.~\eqref{e:D13} associated to the $M_1$ samples that transition from state $\bra j=1\right|$ has two negative entries, i.e., $j=0$ and $j=2$, and one positive entry, i.e. $j=1$. In this case we need to change some of the $M_1$ samples in states $\bra j=0\right|$ and $\bra j=2\right|$ into state $\bra j=1\right|$ according to the ratios specified by these entries (see Fig.~\ref{f:general}). More precisely, following the same reasoning of the previous case, on average we need to change $-M_1\bra 1|\Delta |0\ket = 9\times 2$ out of the $M_1\pi_0 = 9\times 4 $ samples in state $\bra j=0\right|$ and $-M_1\bra 1 |\Delta | 2\ket = 9\times 1$ out of the $M_1\pi_2 = 9\times 5$ samples in state $\bra j=2\right|$ into state $\bra j=1\right|$. This is the reverse process of the previous case: for each sample in state $\bra j=2\right|$ that is changed into state $\bra j=1\right|$, two samples in state $\bra j=0\right|$ are changed into state $\bra j=1\right|$. 

Suppose we save in memory a given number $m_1$ of samples, $\{s_{\ell_k}^t\}_{k=1}^{m_1}$, randomly chosen out of the $M_1$ samples at time step $t$. Then, on average, $m_1\pi_0$ of the corresponding new samples, $\{s_{\ell_k}^{t+1}\}_{k=1}^{m_1}$, generated in the intermediate stage would be in state $\bra j=0\right|$ and $m_1\pi_2$ of them  would be in state $\bra j=2\right|$.  Each of these numbers should be larger than the corresponding number of samples that need to be changed to properly correct the $M_1$ samples, i.e., $m_1\pi_0\geq -M_1\bra 1|\Delta |0\ket$ and $m_1\pi_2\geq -M_1\bra 1|\Delta |2\ket$. The minimum number of samples that need to be saved to typically guarantee these conditions is  
\be
m_1^\ast = \max\left\{-M_1\bra 1|\Delta |0\ket /\pi_0 ,-M_1\bra 1|\Delta |2\ket /\pi_2 \right\},
\ee
 or $m_1^\ast = \tfrac{1}{2}M_1$. Equivalently, the minimum fraction of the $M_1$ samples that we need to save in memory is $f_1 = m_1^\ast / M_1 = \tfrac{1}{2}$---these fractions can be interpreted probabilistically, as discussed in Sec.~\ref{s:gqi}. 

So, if we save in memory only $m_1^\ast$ samples, on average $m_1^\ast\pi_0$ samples will be in state $\bra j=0\right|$ and $m_1^\ast\pi_2$ samples will be in state $\bra j=2\right|$. We need to change $-M_1\bra 1|\Delta |0\ket$ of the $m_1^\ast\pi_0$ samples in state $\bra j=0\right|$ into state $\bra j=1\right|$, or a ratio $r_{1:0\to}^-=-M_1\bra 1|\Delta |0\ket/\left(m_1^\ast\pi_0\right) = 1$. Analgously, we need to change $-M_1\bra 1|\Delta |2\ket$ of the $m_1^\ast\pi_2$ samples in state $\bra j=2\right|$ into state $\bra j=1\right|$, or a ratio $r_{1:2\to}^-=-M_1\bra 1|\Delta |2\ket/\left(m_1^\ast\pi_2\right) = \tfrac{2}{5}$---these ratios can also be interpreted probabilistically, as discussed in Sec.~\ref{s:gqi}. In all these fractions and ratios the actual number of samples $M_1$ cancels out; this is true in general. In the previous case we have $m_0^\ast = -M_0\bra 0|\Delta |1\ket /\pi_1 = \tfrac{1}{3}M_0$, or $f_0 = m_0^\ast / M_0 = \tfrac{1}{3}$, and $r_{0:1\to}^- = 1$; here the actual number of samples $M_0$ also cancels out.

Similarly, the only negative entry in the correction vector $\bra 2\right|\Delta$ in Eq.~\eqref{e:D23} is $\bra 2\right|\Delta\left|1\ket = -{3}/{18}$. To correct the $M_2$ samples corresponding to those that were in state $\bra j=2\right|$ at time step $t$, we need to change on average $-M_2\bra 2\right|\Delta\left|1\ket = 5\times 3$ out of the $M_2\pi_1$ samples that, on average, transitioned into state $\bra j=1\right|$ at time step $t+1$: $M_2\bra 2\right|\Delta\left|0\ket = 5\times 2$ should change, on average, into state $\bra j=0\right|$ and $M_2\bra 2\right|\Delta\left|2\ket = 5\times 1$ should change, on average, into state $\bra j=2\right|$. In this case we have $m_2^\ast = -M_2\bra 2|\Delta |1\ket /\pi_1 = \tfrac{1}{3}M_2$, or $f_2 = m_2^\ast / M_2 = \tfrac{1}{3}$. Additionally, $r^-_{2:1\to} = 1$ since in this case there is only one negative entry. So, all of the corresponding $m_2^\ast\pi_1$ samples that are in state $\bra j=1 \right|$ at the intermediate stage should change. All of these samples should be distributed between states with positive entries in the correction vector $\bra 2\right|\Delta$, i.e., $j=0$ and $j=2$, according to the corresponding ratios. In this case such ratios are $r_{2:\to 0}^+ =\tfrac{1}{Z_2} M_2\bra 2\right|\Delta\left|0\ket = \tfrac{2}{3}$ and $r_{2:\to 2}^+ = \tfrac{1}{Z_2} M_2\bra 2\right|\Delta\left|2\ket = \tfrac{1}{3}$, respectively, where $Z_2$ is a normalization constant enforcing $r_{2:\to 0}^+ + r_{2:\to 2}^+ = 1$. 

Applying the same reasoning to the first case analyzed, i.e., those samples transitioning from state $\bra j=0\right|$, we obtain $r^+_{0:\to 0} = \tfrac{1}{Z_0} M_0\bra 0\right|\Delta\left|0\ket = \tfrac{2}{3}$ and $r^+_{0:\to 2} = \tfrac{1}{Z_0} M_0\bra 0\right|\Delta\left|2\ket = \tfrac{1}{3}$. We obtain the same numbers as before because in this example we have  $\bra 0\right|\Delta = \bra 2\right|\Delta$ (see Eqs.~\eqref{e:D03} and \eqref{e:D23}). This is not true in general, though.

\section{Quantum-like belief propagation protocols}\label{s:bp}

Here we describe the potentially more general perspective discussed in Sec.~\ref{s:sqrt_bp} (Appendix~\ref{s:bp_gen}). Furthermore, we introduce the two graphical models whose belief propagation dynamics is mathematically analogous to the quantum dynamics of the examples discussed in Sec.~\ref{s:qex} (Appendix~\ref{s:bp_ex}). Finally, we point out some caveats that we hope can be resolved in the near future (Appendix~\ref{s:caveats}). 

\subsection{General considerations}\label{s:bp_gen}

\begin{figure*}
\centering
\includegraphics[width=0.98\textwidth]{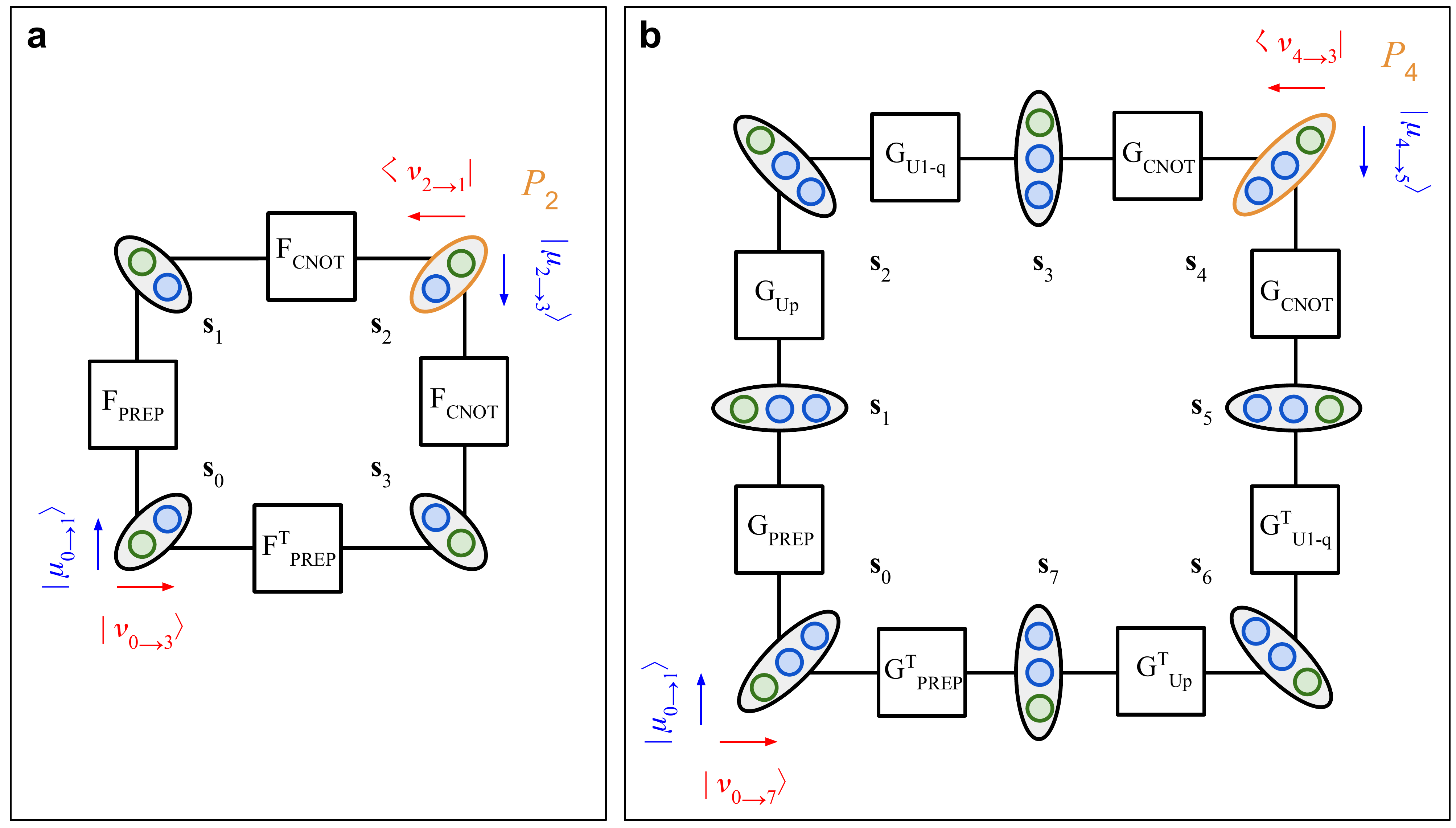}
\caption{{\em Belief propagation as quantum-like dynamics:} Cycle-like graphical models, defined on pairs (a) and triples (b) of binary variables $\bs_\ell$ (ellipses) with $\ell=0,1,\dotsc$, whose belief propagation dynamics is mathematically analogous to the quantum protocols studied in (a) Refs.~\cite{palsson2017experimentally, gu2012quantum} (see Fig.~\ref{f:qis_fig2}a) and in (b) Refs.~\cite{ghafari2018single,thompson2018causal} (see Fig.~\ref{f:qis_fig2}b), respectively. While in Refs.~\cite{realpe2017modeling, realpe2018cognitive}, cycle-like graphical models represent a dynamical interaction between an agent and an experimental device, here we consider these as representing static Ising-like systems. Factors (squares) represent interactions between `consecutive' pairs of binary variables (see Eqs.~\eqref{e:Fx},\eqref{e:Fprep2}, and \eqref{e:Fprep}-\eqref{e:Fcnot} for the definition of all factors in (a) and (b)). Although belief propagation is not guaranteed to be exact on graphs with loops, if we choose `initial' messages such that the binary variables in $\bs_0$ are all zero with probability one, belief propagation becomes exact (see Sec.~\ref{s:bp}). Messages (a) $\left|\mu_{2\to 3}\ket$ and (b) $\left|\mu_{4\to 5}\ket$ have the same mathematical form as the quantum states in (a) Eqs.~\eqref{e:S1}-\eqref{e:S2} and (b) Eqs.~\eqref{e:|0'>}-\eqref{e:|2'>}, respectively, depending on the choice of factors $F_{\rm PREP}$ and $G_{\rm PREP}$. Furthermore, messages $\bra\nu_{\ell\to \ell -1}\right| = \left[\left|\mu_{\ell\to \ell +1}\ket \right]^\dag$ can be obtained via the hermitian conjugate operation, as in quantum theory.}\label{f:qis_fig3}
\end{figure*}

Belief propagation is a message-passing algorithm to efficiently compute marginals of graphical models by passing messages from node to node along the underlying graph (see Fig.~\ref{f:qis_fig3}; see also Sec.~14 in Ref.~\cite{mezard2009information}). It was recently shown~\cite{realpe2017modeling,realpe2018cognitive} (see Secs.~V~B and C in Ref.~\cite{realpe2017modeling}) that belief propagation on chain- or cycle-like graphs is mathematically analogous to quantum dynamics in imaginary time, i.e., the dynamics obtained after changing time $t$ to $-it$, where $i$ is the imaginary unit. Since the imaginary unit only appears multiplying the phase of a wave function, it is natural to expect that quantum protocols that do not exploit information about the phase, like those presented in Sec.~\ref{s:quantum} could be efficiently simulated classically.

For instance, let us consider a graphical model of $N+1$ binary variables $s_\ell\in\{0,1\}$, with $\ell=0,\dotsc , N$, interacting on a cycle whose probability distribution factorizes as (see Fig.~\ref{f:qis_fig3}) 
\be\label{e:GM}
\mathcal{P}(s_0,\dotsc, s_N) = \frac{1}{Z}\prod_{\ell=0}^{N}F_\ell(s_{\ell +1} ,s_{\ell }), 
\ee
where periodicity in the index $\ell$ is understood, i.e., $s_{N+1}=s_0$ Here the non-negative functions $F_\ell(s_{\ell +1}, s_{\ell})$, for $\ell =0,\dotsc N$, are called factors and $Z$ is the normalization constant. For instance, if $\mathcal{P}$ is a Boltzmann distribution, the factor $F_\ell$ can be a Boltzmann weight $F_\ell(s_{\ell +1}, s_\ell) = e^{-\beta E_\ell(s_{\ell+1} , s_\ell)}$ associated to an energy function $E_\ell$ and an inverse temperature $\beta$. It is straightforward to extend the ideas presented here to $D$-dimensional arrays of binary variables $\bs_\ell = (s_\ell^{(1)},\dotsc ,s_\ell^{(D)} )$ (see Sec.~\ref{s:bp_ex} and Fig.~\ref{f:qis_fig3}).

Belief propagation is guaranteed to yield exact marginals when the graphical model has the topology of a tree. So, let us first assume that factor $F_N(s_0,s_N)=1$ for all $s_0, s_N\in\{0,1\}$, so the cycle turns effectively into a chain. In this case, belief propagation passes messages $\mu_{\ell\to\ell +1}(s_\ell)$ and $\nu_{\ell\to \ell -1}(s_\ell)$ starting at nodes $\ell =0$ and $\ell =N$ respectively~\cite{mezard2009information} (see Sec.~14 therein). After such messages have travelled the entire chain, using a suitable normalization for the messages, the marginal probability of variable $s_\ell$ is given by~\cite{realpe2017modeling,realpe2018cognitive} (see Sec.~V~B in Ref.~\cite{realpe2017modeling}) 
\be\label{e:marginal}
p_\ell(s_\ell) = \mu_{\ell\to\ell +1}(s_\ell)\nu_{\ell\to\ell -1}(s_\ell).
\ee
This is reminiscent of the Born rule of quantum theory, and we can indeed write these messages as~\cite{realpe2017modeling, realpe2018cognitive} 
\BE
\mu_{\ell\to\ell +1}(s) &=& \sqrt{p_\ell(s)}e^{\phi_{\ell} (s)},\label{e:mu}\\
\nu_{\ell\to\ell -1}(s) &=& \sqrt{p_\ell(s)}e^{-\phi_\ell (s)}.\label{e:nu}
\EE

If we do $t \to -i t$ in the Schr\"odinger equation, the imaginary unit $i$ disapears. The imaginary unit multiplying the phase $\varphi_\ell(s)$ of a wave function $\psi_\ell(s) = \sqrt{p_\ell(s)}e^{i\varphi_\ell(s)}$ also disappears, which leads to the imaginary-time analogs of a wave function and its conjugate, Eqs.~\eqref{e:mu} and \eqref{e:nu}. Furthermore, if we write 
\BE
\left|\mu_{\ell\to\ell +1}\ket &=& \begin{pmatrix} \mu_{\ell\to\ell +1}(0)\\ \mu_{\ell\to\ell +1}(1)\end{pmatrix} ,\\
\bra\nu_{\ell \to\ell -1}\right| &=& \begin{pmatrix} \nu_{\ell\to\ell -1}(0), & \nu_{\ell\to\ell -1}(1)\end{pmatrix} ,
\EE
the belief propagation algorithm associated to Eq.~\eqref{e:GM} is determined by the update rules 
\BE
\left|\mu_{\ell\to\ell +1}\ket &=& F_{\ell -1}\left|\mu_{\ell -1\to\ell }\ket,\label{e:BP->}\\
\bra\nu_{\ell \to\ell -1}\right| &=& \bra\nu_{\ell +1\to\ell }\right| F_{\ell },\label{e:BP<-}
\EE
which are similar to the update rules of quantum theory, except that the factors $F_\ell$ are real and non-negative. However, see Refs.~\cite{realpe2017modeling, realpe2018cognitive} for a discussion of more general situations. 

From a mathematical point of view, the only difference between Eqs.~\eqref{e:mu} and \eqref{e:nu} and quantum wavefunctions is the lack of the imaginary unit $i$ multiplying the phase $\phi_\ell$. So it is natural to expect that quantum protocols that only exploit amplitudes of wave functions, such as those based in Eqs.~\eqref{e:S1}, \eqref{e:S2} and \eqref{e:|0'>}-\eqref{e:|2'>}, might be implemented classically, e.g., via belief propagation algorithms. This is not totally unexpected given that $\sqrt{p_\ell(s_\ell)}$ contains the same information encoded in $p_\ell(s_\ell)$. We will argue here that this may indeed be the case in some instances.

\subsection{Examples}\label{s:bp_ex}

\subsubsection{Symmetrically perturbed coin process}\label{s:bp_1}
To gain some initial intuition, we will first use an alternative method also introduced in Refs.~\cite{realpe2017modeling,realpe2018cognitive} (see Sec.~VI~C in Ref.~\cite{realpe2017modeling}), which, instead of messages, uses probability matrices that are analogous to the imaginary-time version of density matrices. Due to the absence of phases in the examples considered here, such probability matrices actually have the same mathematical form as the corresponding density matrices of interest, as we will see below. 

Figure~\ref{f:qis_fig3}a shows a graphical model, defined on pairs of binary variables $\bs_\ell = (s_\ell^{(1)},s_\ell^{(2)})$ with $\ell =0,\dotsc ,3$, whose belief propagation dynamics is mathematically analogous to the quantum protocol of the symmetrically perturbed coin process described in Sec.~\ref{s:qex}. As we mentioned in Sec.~\ref{s:qex}, the undetermined entries in Eq.~\eqref{e:Ux} are irrelevant for the application of the quantum protocols described therein. They are chosen such that $U_x$ is unitary, which requires some entries to be negative. However, as they are irrelevant, so we can also choose them equal to zero in the belief propagation model, for instance. We can then implement the corresponding transformations using a classical factor 
\be\label{e:Fx}
F_x = \begin{pmatrix} \sqrt{1-x} & 0 \\ \sqrt{x} & 0\end{pmatrix}
\ee
instead (cf.~Eq.~\eqref{e:Ux}).

Using Eq.~\eqref{e:Fx}, the factors in Fig.~\ref{f:qis_fig3}a can be defined as
\be\label{e:Fprep2}
F_{\rm PREP} = F_{x_j}\otimes F_p\hspace{0.3cm}\textrm{ and } \hspace{0.3cm} F_{\rm CNOT} = \mathrm{CNOT}^{(1,2)},
\ee

\

\noindent where $\mathrm{CNOT}^{(1,2)}$ is defined in Eq.~\eqref{e:CNOT2}, and $j=0,1$ refers to the causal states. In this case, $x_0=p$ and $x_1=1-p$, to implement the analogs of quantum causal states $\left|\xi_0\ket$ and $\left|\xi_1\ket$, respectively.  

The probability of a path $\bs = (\bs_0 , \dotsc , \bs_3)$ can be written as~\cite{realpe2017modeling,realpe2018cognitive}
\begin{widetext}
\be
\mathcal{P}(\bs) = F^T_{\rm PREP}(\bs_0^\prime ,\bs_3) F^T_{\rm CNOT}(\bs_3 ,\bs_2) F_{\rm CNOT}(\bs_2 ,\bs_1) F_{\rm PREP}(\bs_1 ,\bs_0),\label{e:prob}
\ee
\end{widetext}
where we have written $\bs_0^\prime = \bs_0$ for future convenience---here the normalization constant is $Z=1$. Due to the circular topology of the graphical model in Fig.~\ref{f:qis_fig3}a it is in general not possble to compute the marginal $p_{1}$ at step $\ell =1$, for instance, from the marginal
\be\label{e:Pmatrix}
p_{0}(\bs_0) = \sum_{\bs_1,\bs_2,\bs_3}\mathcal{P}(\bs),
\ee
at step $\ell=0$ alone~\cite{realpe2018cognitive}. If the graphical model had the topology of a chain, instead, this would indeed be possible, i.e., we would have a Markov chain. 

However, if we relax the condition that $\bs_0^\prime = \bs_0$ in Eqs.~\eqref{e:prob} and\eqref{e:Pmatrix}, and interpret the factors as probability matrices, we can define a real {\em probability matrix}~\cite{realpe2017modeling,realpe2018cognitive} (see Sec. VI C in Ref.~\cite{realpe2017modeling})
\be\label{e:P0}
P_{0} = F^T_{\rm PREP} F^T_{\rm CNOT} F_{\rm CNOT}F_{\rm PREP}= \left|0\ket\bra 0\right|\otimes\left|0\ket\bra 0\right| ,
\ee
at step $\ell =0$. The $\bs_0$-th diagonal element of $P_0$ in Eq.~\eqref{e:P0} is the marginal $p_0(\bs_0)$, which in this case equals one if $\bs_0=(0,0)$ and zero otherwise. 

Following the same reasoning, and taking into account that $F_{\rm CNOT}F_{\rm CNOT}=\id$, we can compute a probability matrix (see Eq.~\eqref{e:Fprep2})
\be\label{e:P1}
P_{1} = F_{\rm PREP}F^T_{\rm PREP} F_{\rm CNOT} F_{\rm CNOT}= \left|\xi_j\ket\bra \xi_j\right| \otimes\left|\xi_0\ket\bra \xi_0\right| ,
\ee
at step $\ell =1$. Again, the diagonal of $P_1$ contains the vector of marginal probabilities $p_1$. In this case, the matrix $P_1$ is mathematically analogous to the density matrix associated to the quantum causal state $\left|\xi_j\ket\left|\xi_0\ket$ of the quantum protocol shown in Fig.~\ref{f:qis_fig2}b (see Eqs.~\eqref{e:S1}, \eqref{e:S2}, and \eqref{e:chi}).

Matrix $P_1$ in Eq.~\eqref{e:P1} is obtained by the cyclic permutation of the final matrix $F_{\rm PREP}$ in Eq.~\eqref{e:P0}. Continuing with these cyclic permutations, we get (see Eqs.~\eqref{e:CNOT|S1>|S1>} and \eqref{e:CNOT|S2>|S1>})
\BE
P_{2} &=& \left|\chi_j\ket\bra\chi_j\right|,\label{e:P2} \\
P_{3} &=& P_{1},
\EE
after which we get back to $P_0 = \left|0\ket\bra 0\right|\otimes \left|0\ket\bra 0\right|$, so the dynamics is self-consistent. The probability matrix $P_2$ has the same mathematical form of the density matrix associated to states $\left|\chi_j\ket$, with $j\in\{0,1\}$, in Eqs.~\eqref{e:CNOT|S1>|S1>} and \eqref{e:CNOT|S2>|S1>}.

By measuring the first binary variable of $P_2$ we get the same statistics and the same state update for the probabilistic state of the second variable as the quantum protocol in Fig.~\ref{f:qis_fig2}b.

Now, let us turn back to the belief propagation algorithm, Eqs.~\eqref{e:BP->} and \eqref{e:BP<-}, which, again, do not in general yield exact marginals for graphical models with loops. In this case, however, the marginals estimated by belief propagation (see Eq.~\eqref{e:marginal}) are indeed exact if we pick ``initial'' messages (i.e., at index $\ell=0$)~\cite{realpe2017modeling,realpe2018cognitive} 
\BE
\left|\mu_{0\to 1}\ket &=& \left|0\ket\left|0\ket ,\\
\bra\nu_{0\to 3}\right| &=& \bra 0 \right|\bra 0\right| ,
\EE
which are consistent with $P_0$ in Eq.~\eqref{e:P0}. 

The forward belief propagation iteration, Eq.~\eqref{e:BP->}, leads in this case to (see Fig.~\ref{f:qis_fig3}a and Eqs.~\eqref{e:S1},\eqref{e:S2}, \eqref{e:CNOT|S1>|S1>} and \eqref{e:CNOT|S2>|S1>})
\BE
\left|\mu_{1\to 2}\ket &=& F_{\rm PREP}\left|\mu_{0\to 1}\ket = \left|\xi_j\ket\left|\xi_0\ket,\\
\left|\mu_{2\to 3}\ket &=& F_{\rm CNOT} \left|\mu_{1\to 2}\ket = \left|\chi_j\ket ,\label{e:mu2->3} \\ 
\left|\mu_{3\to 0}\ket &=& F_{\rm CNOT} \left|\mu_{2\to 3}\ket = \left|\mu_{1\to 2}\ket \label{e:mu3->0}, \\
\left|\mu_{0\to 1}\ket &=& F^{T}_{\rm PREP}\left|\mu_{3\to 0}\ket =\left|\mu_{0\to 1}\ket.\label{e:mu0->1}
\EE
The message $\left|\mu_{1\to 2}\ket$ at $\ell=1$ is mathematically equivalent to the initial quantum state $\left|\xi_j\ket\left|\xi_0\ket$ of the quantum protocol (see Fig.~\ref{f:qis_fig2}b and Eqs.~\eqref{e:S1} and \eqref{e:S2}). Similarly, the message $\left|\mu_{2\to 3}\ket$ at $\ell=2$ is mathematically equivalent to the quantum state \eqref{e:chi}. Finally, after the last two steps, Eqs.~\eqref{e:mu3->0} and \eqref{e:mu0->1}, we turn around the cycle and belief propagation consistently yields back the initial message $\left|\mu_{0\to 1}\ket $. So the forward belief propagation iteration, Eq.~\eqref{e:BP->}, is self-consistent, even though the graphical model does not have the topology of a tree.

The backward belief propagation iteration, Eq.~\eqref{e:BP<-}, yields the transposed equations, i.e., at each time index $\ell$ we have 
\be
\bra\nu_{\ell\to \ell-1}\right| = \left[\left|\mu_{\ell\to \ell+1}\ket \right]^{T},
\ee
which is equivalent to taking the hermitian conjugate, as in quantum theory, since the messages are real. 

\

\subsubsection{Post-processed perturbed coin process}
Figure~\ref{f:qis_fig3}b shows a graphical model, defined on binary variables $\bs_\ell = (s_\ell^{(1)},s_\ell^{(2)},s_\ell^{(3)})$ with $\ell=0,\dotsc ,7$, where belief propagation follows a dynamics analogous to that of the quantum protocol for the post-processed perturbed coin process in Sec.~\ref{s:qex} (see Fig.~\ref{f:qis_fig2}b). The factors in Fig.~\ref{f:qis_fig3}b are defined as
\BE
G_{\rm PREP} &=& G_j\otimes\left|0\ket\bra 0\right| \otimes\left|0\ket\bra 0\right|,\label{e:Fprep}\\ 
G_{\rm U_p} &=& \left|0\ket\bra 0\right| \otimes\id \otimes F_p + \left|1\ket\bra 1\right| \otimes\id\otimes\id,\label{e:-CFp} \\ 
G_{\rm U_{1-q}} &=& \left|0\ket\bra 0\right| \otimes\id \otimes \id + \left|1\ket\bra 1\right| \otimes F_{1-q}\otimes\id, \label{e:CFq}\\
G_{\rm CNOT} & =& \mathrm{CNOT}^{(3,2)},\label{e:Fcnot}
\EE
where $F_p$ and $F_{1-q}$ are obtained by setting $x=p$ and $x=1-q$ in Eq.~\eqref{e:Fx}, respectively, and $\mathrm{CNOT}^{(3,2)}$ is defined in Eq.~\eqref{e:CNOT}. Factors $G_{\rm Up}$ and $G_{\rm U_{1-q}}$ are analogous to the gates in Eqs.~\eqref{e:-CUp} and \eqref{e:CUq}. The factor $G_j$ prepares from $\left|0\ket$ the analog of the initial quantum causal state $\left|\xi_j\ket$, with $j=0,1,2$. More precisely, $G_0 = \left| 0\ket\bra 0\right|$, or $G_1 = F_q$, or $G_2= \left| 1\ket\bra 0\right|$ if the initial quantum causal state is $\left|\xi_0\ket$, or $\left|\xi_1\ket$, or $\left|\xi_2\ket$, respectively  (see Eqs.~\eqref{e:|0'>}-\eqref{e:|2'>}).

As before, the graphical model in Fig.~\ref{f:qis_fig3}b has the topology of a circle and belief propagation is not guaranteed in general to yield exact marginals for graphical models with cycles. However, the variable $\bs_0$ is connected to factors $G_{\rm PREP}$ and $G_{\rm PREP}^T$, whose entries in the second column and second row, respectively, are zero for any $j=0,1,2$ (see Eq.~\eqref{e:Fprep}). This implies that $\bs_0 = (0,0,0)$ with probability one---to see this notice that 
\be\label{e:|000>}
G_{\rm PREP}^T G_{\rm PREP} = \left|000\ket\bra 00 0\right|. 
\ee
In this case, the message-passing equations become exact if, consistent with Eq.~\eqref{e:|000>}, we pick ``initial'' messages (i.e., at index $\ell=0$)~\cite{realpe2017modeling,realpe2018cognitive} (see Sec. V B 2 in Ref.~\cite{realpe2017modeling}) 
\BE
\left|\mu_{0\to 1}\ket &=& \left|0\ket\left|0\ket\left|0\ket\label{e:mu3_0->}  ,\\
\bra\nu_{0\to 7}\right| &=& \bra 0 \right|\bra 0 \right|\bra 0\right| .\label{e:nu3_0<-}
\EE

The first iteration of the belief propagation equations, Eqs.~\eqref{e:BP->} and \eqref{e:BP<-}, yields 
\BE
\left|\mu_{1\to 2}\ket &=& G_{\rm PREP}\left|\mu_{0\to 1}\ket = \left|\xi_j\ket\left|0\ket\left|0\ket,\\
\bra\nu_{7\to 6}\right|&=& \bra\nu_{0\to 7}\right|G_{\rm PREP}^T =\bra \xi_j \right|\bra 0 \right|\bra 0\right|,
\EE
where $\left|\xi_j\ket = F_j\left|0\ket$, which has the same mathematical form as the initial quantum causal state of the protocol shown in Fig.~\ref{f:qis_fig3}b (see Eqs.~\eqref{e:|0'>}-\eqref{e:|2'>}).

Now $F_p$ and $F_{1-q}$ coincide, respectively, with $U_p$ and $U_{1-q}$, except in the undetermined entries $\#$ of the latter (see Eqs.~\eqref{e:Fx} and \eqref{e:Ux}). However, such undetermined entries are irrelevant for the quantum protocol in Fig.~\ref{f:qis_fig2}b (see Sec.~\ref{s:qex}). As far as this quantum protocol is concerned, Eqs.~\eqref{e:-CFp}-\eqref{e:Fcnot} are equivalent to Eqs.~\eqref{e:-CUp}-\eqref{e:CNOT}. This implies that the belief propagation dynamics induced on the messages $\left|\mu_{\ell\to\ell +1}\ket$ and $\bra\nu_{\ell\to\ell -1}\right|$ by Eqs.~\eqref{e:BP->} and \eqref{e:BP<-} with factors $G_{\rm Up}$, $G_{\rm U_{1-q}}$, and $G_{\rm CNOT}$ defined in Eqs.~\eqref{e:-CFp}-\eqref{e:Fcnot}, has the same mathematical form as the dynamics of the corresponding quantum protocol in Fig.~\ref{f:qis_fig2}b. In particular (see Eqs.~\eqref{e:U0}-\eqref{e:U2}), 
\be
\left|\mu_{4\to 5}\ket = U\left|\xi_j\ket\left|0\ket\left|0\ket ,\label{e:mu3->}
\ee
which yields the same statistics as the post-processed perturbed coin process (see Fig.~\ref{f:qis_fig1}b), as described in Sec.~\ref{s:qex}. 

The branch of the graphical model in Fig.~\ref{f:qis_fig3}a that connects variables $\bs_\ell$, with $\ell=0,1,2,3,4$, ``prepares'' the message in Eq.~\eqref{e:mu3->}. The remaining branch connecting variables $\bs_\ell$, with $\ell=0,7,6,5,4$, is a kind of ``mirror image'' of the former, i.e., it has the same factors but transposed (see Fig.~\ref{f:qis_fig3}a)---such mirror image yields the analogs of the corresponding conjugate wave functions. Indeed, since the initial $\nu$-message in Eq.~\eqref{e:nu3_0<-} is also the transpose of the initial $\mu$-message in Eq.~\eqref{e:mu3_0->}, the dynamics of the $\nu$-messages is the same as that of the $\mu$-messages. So, the $\nu$-messages have the same mathematical form of the conjugate wave functions in the corresponding quantum protocol (see Fig.~\ref{f:qis_fig2}a and Example 1 in Sec.~\ref{s:qex}). This happens because there are no phases involved.

\begin{figure*}
\centering
\includegraphics[width=0.7\textwidth]{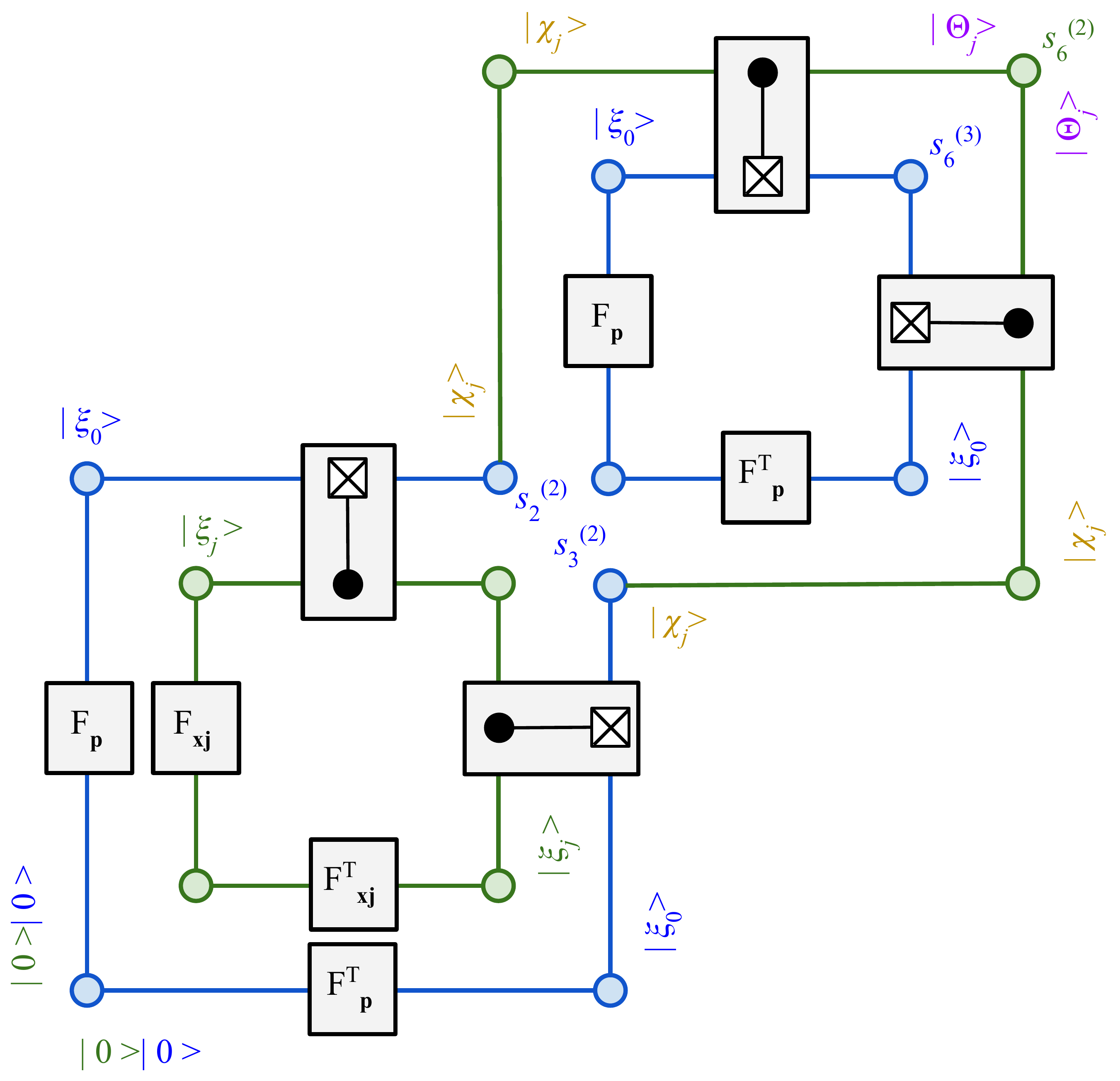}
\caption{{\em Two-step belief propagation protocol:} Extension of the graphical model in Fig.~\ref{f:qis_fig3}a whose belief propagation dynamics is mathematically similar to two iterations of the corresponding quantum protocol in Fig.~\ref{f:qis_fig2}a. Here we show explicitly all variables and factors involved, which are defined in Eq.~\eqref{e:Fprep2}. Factor $F_{\rm CNOT}$ is here represented with a notation similar to the corresponding ${\rm CNOT}^{(1,2)}$ gate in Fig.~\ref{f:qis_fig2}a since it is assymetric in the variables $s^{(1)}_\ell$ and $s^{(2)}_\ell$. This graphical model is essentially composed of two copies of the graphical model in Fig.~\ref{f:qis_fig3}a connected through variables $s_2^{(1)}$ and $s_2^{(2)}$, except that the second copy (top right) does not need the preparation factor $F_{x_j}$, similar to the case of the quantum protocol---otherwise we would have to re-prepare the state at each iteration destroying any potential memory savings. Messages $\left|\xi_j\ket$ and $\left|\chi_j\ket$ have the same mathematial form of the quantum states in Eqs.~\eqref{e:S1}-\eqref{e:S2} and Eqs.~\eqref{e:CNOT|S1>|S1>}-\eqref{e:CNOT|S2>|S1>}, respectively. Messages $\left|\Theta_j\ket = {\rm CNOT}^{(2,3)}\left|\chi_j\ket\left|\xi_0\ket$ correspond to a second iteration of the quantum protocol in Fig.~\ref{f:qis_fig2}a. We can generate further iterations by opening the (blue) top right variable $s_6^{(3)}$ and repeating the process. If we interpret this graphical model as a static Ising-like system, instead of requiring less memory to generate samples than the corresponding classical $\epsilon$-machine we would rather have a proliferation of stochastic bits to implement the whole circuit. A potential way out is to implement each factor at a time, e.g., between variables $\bs_0$ and $\bs_1$, remove those factors along with variable $\bs_0$, then implement the next factor between $\bs_1$ and $\bs_2$, and so on. This is closer to the original dynamical interpretation in Refs.~\cite{realpe2017modeling,realpe2018cognitive} which, being dynamical, in principle completely avoids this problem. However, this may require significant isolation of the system from the environment.
}\label{f:bp2}
\end{figure*}

Finally, the remaining iterations of belief propagation for the $\mu$-messages roll back the messages. For instance, since $G_{\rm CNOT}G_{\rm CNOT}= \id$, we get
\be
\left|\mu_{5\to 6}\ket = \left|\mu_{3\to 4}\ket = G \left|\mu_{0\to 1}\ket = G \left|0\ket,
\ee
with 
\begin{widetext}
\be
G= G_{\rm U_{1-q}} G_{\rm U_p} G_{\rm PREP} = \left| 0 \ket\bra 0\right| G_j\otimes \left| 0 \ket\bra 0\right|\otimes F_p\left| 0 \ket\bra 0\right|+\left| 1 \ket\bra 1\right| G_j\otimes F_{1-q}\left| 0 \ket\bra 0\right|\otimes \left| 0 \ket\bra 0\right|
\ee
\end{widetext}
Now, since $F_x^T F_x = \left| 0\ket\bra 0\right|$, we get $G^T G = \left|000\ket\bra 000\right|$ for all $j=0,1,2$. So, after turning around the cycle we get
\be
G^T G \left| 0\ket = \left| 0\ket =\left|\mu_{0\to 1}\ket , 
\ee
which shows the belief propagation equations are self-consistent in this case too (cf. Sec.~\ref{s:bp_1}).

\subsection{Quantum-like protocols and some caveats}\label{s:caveats}
The graphical models introduced in Sec.~\ref{s:bp_ex} have a belief propagation dynamics mathematically analogous to the first iteration of the quantum protocols in Fig.~\ref{f:qis_fig2}. This first iteration includes the preparation of the message $\left|\xi_j\ket$ via factor $F_{x_j}$. However, this is to be avoided in future iterations---otherwise any potential memory savings would be destroyed by having to prepare message $\left|\xi_j\ket$ at each time step. 

Figure~\ref{f:bp2} shows an extension of the graphical model in Fig.~\ref{f:qis_fig3}a whose dynamics is mathematically analogous to that of two iterations of the quantum protocol in Fig.~\ref{f:qis_fig2}a. This is essentially composed of two copies of the graphical model in Fig.~\ref{f:qis_fig3}a connected through the variables $s_2^{(2)}$ and $s_3^{(2)}$ which, in contrast to variable $s_2^{(1)}$, are not observed. The second copy (top right) does not have the preparation factor $F_{x_j}$ as required. This construction can be continued iteratively by splitting variable $s_6^{(3)}$ into two and connecting a third copy to them, and so on. A similar construction can be done for the graphical model in Fig.~\ref{f:qis_fig3}b. There are some caveats, though.

First, if we interpret the graphical model in Fig.~\ref{f:bp2} as a static Ising-like system, instead of reducing the amount of memory required to generate stochastic trajectories, in comparison to the corresponding $\epsilon$-machine, the number of stochastic bits needed to implement the whole graphical model would rather proliferate. A potential way out is to implement each factor graph at a time (see Fig.~\ref{f:bp2}). This would be closer to the original interpretation of such type of graphical models in Refs.~\cite{realpe2017modeling,realpe2018cognitive} as describing the dynamics of a physical agent, e.g., a robot, interacting with an experimental device. However, this may require significant isolation from the environment to avoid the probabilities, and messages, associated to such isolated variables from changing too much. Whether the amount of isolation required is similar to that of quantum protocols is an open question at the moment.

Second, the memory gain in the case of the symmetrically perturbed coin process is in terms of statistical memory, which needs a quantum-like encoding \`a la Schumacher~\cite{schumacher1995quantum}. In the example studied here such an encoding amounts essentially at working with the Fourier transform of the messages. It is not clear at this point how to implement this either physically or algorithmically. This caveat does not apply for the topological memory savings in the case of the belief propagation protocol for the post-processed perturbed coin process. However, in more general cases a similar caveat may arise also for topologically memory saving graphical models if the lower-dimensional representation of the messages corresponding to the quantum causal states (see Eqs.~\eqref{e:|0'>}-\eqref{e:|2'>}) involves negative numbers. This may be dealt with using the operational interpretation~\cite{realpe2018cognitive} of this type of negative numbers we have discussed in Secs.~\ref{s:c-examples} and \ref{s:extensions}.


\begin{thebibliography}{10}

\bibitem{bouchaud2018trades}
J.-P. Bouchaud, J.~Bonart, J.~Donier, and M.~Gould, {\em Trades, quotes and
  prices: financial markets under the microscope}.
\newblock Cambridge University Press, 2018.

\bibitem{bisias2012survey}
D.~Bisias, M.~Flood, A.~W. Lo, and S.~Valavanis, ``A survey of systemic risk
  analytics,'' {\em Annu. Rev. Financ. Econ.}, vol.~4, no.~1, pp.~255--296,
  2012.

\bibitem{farmer2009economy}
J.~D. Farmer and D.~Foley, ``The economy needs agent-based modelling,'' {\em
  Nature}, vol.~460, no.~7256, p.~685, 2009.

\bibitem{stern2016economics}
N.~Stern, ``Economics: current climate models are grossly misleading,'' {\em
  Nature News}, vol.~530, no.~7591, p.~407, 2016.

\bibitem{cai2015environmental}
Y.~Cai, K.~L. Judd, T.~M. Lenton, T.~S. Lontzek, and D.~Narita, ``Environmental
  tipping points significantly affect the cost- benefit assessment of climate
  policies,'' {\em Proceedings of the National Academy of Sciences},
  p.~201503890, 2015.

\bibitem{franzke2015stochastic}
C.~L. Franzke, T.~J. O'Kane, J.~Berner, P.~D. Williams, and V.~Lucarini,
  ``Stochastic climate theory and modeling,'' {\em Wiley Interdisciplinary
  Reviews: Climate Change}, vol.~6, no.~1, pp.~63--78, 2015.

\bibitem{lake2015human}
B.~M. Lake, R.~Salakhutdinov, and J.~B. Tenenbaum, ``Human-level concept
  learning through probabilistic program induction,'' {\em Science}, vol.~350,
  no.~6266, pp.~1332--1338, 2015.

\bibitem{ghahramani2015probabilistic}
Z.~Ghahramani, ``Probabilistic machine learning and artificial intelligence,''
  {\em Nature}, vol.~521, no.~7553, p.~452, 2015.

\bibitem{goodfellow2016deep}
I.~Goodfellow, Y.~Bengio, and A.~Courville, {\em Deep Learning}.
\newblock MIT Press, 2016.
\newblock \url{http://www.deeplearningbook.org}.

\bibitem{murphy2012machine}
K.~P. Murphy, {\em Machine learning: a probabilistic perspective}.
\newblock MIT press, 2012.

\bibitem{mohseni2017commercialize}
M.~Mohseni, P.~Read, H.~Neven, S.~Boixo, V.~Denchev, R.~Babbush, A.~Fowler,
  V.~Smelyanskiy, and J.~Martinis, ``Commercialize quantum technologies in five
  years,'' {\em Nature News}, vol.~543, no.~7644, p.~171, 2017.

\bibitem{boixo2018characterizing}
S.~Boixo, S.~V. Isakov, V.~N. Smelyanskiy, R.~Babbush, N.~Ding, Z.~Jiang, M.~J.
  Bremner, J.~M. Martinis, and H.~Neven, ``Characterizing quantum supremacy in
  near-term devices,'' {\em Nature Physics}, vol.~14, no.~6, p.~595, 2018.

\bibitem{biswas2017nasa}
R.~Biswas, Z.~Jiang, K.~Kechezhi, S.~Knysh, S.~Mandra, B.~O’Gorman,
  A.~Perdomo-Ortiz, A.~Petukhov, J.~Realpe-G{\'o}mez, E.~Rieffel, {\em et~al.},
  ``{A NASA perspective on quantum computing: Opportunities and challenges},''
  {\em Parallel Computing}, vol.~64, pp.~81--98, 2017.

\bibitem{gu2012quantum}
M.~Gu, K.~Wiesner, E.~Rieper, and V.~Vedral, ``Quantum mechanics can reduce the
  complexity of classical models,'' {\em Nature Communications}, vol.~3,
  p.~762, 2012.

\bibitem{thompson2018causal}
J.~Thompson, A.~J. Garner, J.~R. Mahoney, J.~P. Crutchfield, V.~Vedral, and
  M.~Gu, ``Causal asymmetry in a quantum world,'' {\em Physical Review X},
  vol.~8, no.~3, p.~031013, 2018.

\bibitem{aghamohammadi2018extreme}
C.~Aghamohammadi, S.~P. Loomis, J.~R. Mahoney, and J.~P. Crutchfield, ``Extreme
  quantum memory advantage for rare-event sampling,'' {\em Physical Review X},
  vol.~8, no.~1, p.~011025, 2018.

\bibitem{ghafari2018single}
F.~Ghafari, N.~Tischler, J.~Thompson, M.~Gu, L.~K. Shalm, V.~B. Verma, S.~W.
  Nam, R.~B. Patel, H.~M. Wiseman, and G.~J. Pryde, ``Single-shot quantum
  memory advantage in the simulation of stochastic processes,'' {\em arXiv
  preprint arXiv:1812.04251}, 2018.

\bibitem{ghafari2019interfering}
F.~Ghafari, N.~Tischler, C.~Di~Franco, J.~Thompson, M.~Gu, and G.~J. Pryde,
  ``Interfering trajectories in experimental quantum-enhanced stochastic
  simulation,'' {\em Nature {C}ommunications}, vol.~10, no.~1, p.~1630, 2019.

\bibitem{palsson2017experimentally}
M.~S. Palsson, M.~Gu, J.~Ho, H.~M. Wiseman, and G.~J. Pryde, ``Experimentally
  modeling stochastic processes with less memory by the use of a quantum
  processor,'' {\em Science Advances}, vol.~3, no.~2, p.~e1601302, 2017.

\bibitem{binder2018practical}
F.~C. Binder, J.~Thompson, and M.~Gu, ``Practical unitary simulator for
  non-{M}arkovian complex processes,'' {\em Physical Review Letters}, vol.~120,
  no.~24, p.~240502, 2018.

\bibitem{liu2018optimal}
Q.~Liu, T.~Elliott, F.~Binder, C.~Di~Franco, M.~Gu, {\em et~al.}, ``Optimal
  stochastic modelling with unitary quantum dynamics,'' {\em arXiv preprint
  arXiv:1810.09668}, 2018.

\bibitem{thompson2017using}
J.~Thompson, A.~Garner, V.~Vedral, M.~Gu, {\em et~al.}, ``Using
  quantum theory to simplify input--output processes,'' {\em npj Quantum
  Information}, vol.~3, no.~1, p.~6,  2017.

\bibitem{crutchfield1989inferring}
J.~P. Crutchfield and K.~Young, ``Inferring statistical complexity,'' {\em
  Physical Review Letters}, vol.~63, no.~2, p.~105, 1989.

\bibitem{shalizi2001computational}
C.~R. Shalizi and J.~P. Crutchfield, ``Computational mechanics: Pattern and
  prediction, structure and simplicity,'' {\em Journal of Statistical Physics},
  vol.~104, no.~3-4, pp.~817--879, 2001.

\bibitem{schumacher1995quantum}
B.~Schumacher, ``Quantum coding,'' {\em Physical Review A}, vol.~51, no.~4,
  p.~2738, 1995.

\bibitem{wiebe2014quantum}
N.~Wiebe, A.~Kapoor, and K.~M. Svore, ``Quantum deep learning,'' {\em arXiv
  preprint arXiv:1412.3489}, 2014.

\bibitem{wiebe2015quantum}
N.~Wiebe, A.~Kapoor, C.~Granade, and K.~M. Svore, ``Quantum inspired training
  for {B}oltzmann machines,'' {\em arXiv preprint arXiv:1507.02642}, 2015.

\bibitem{tang2018quantum}
E.~Tang, ``A quantum-inspired classical algorithm for recommendation systems,''
  {\em arXiv preprint arXiv:1807.04271}, 2018.

\bibitem{gilyen2018quantum}
A.~Gily{\'e}n, S.~Lloyd, and E.~Tang, ``Quantum-inspired low-rank stochastic
  regression with logarithmic dependence on the dimension,'' {\em arXiv
  preprint arXiv:1811.04909}, 2018.

\bibitem{hen2018quantum}
I.~Hen, ``How quantum is the speedup in adiabatic unstructured search?,'' {\em
  arXiv preprint arXiv:1811.08302}, 2018.

\bibitem{arrazola2019quantum}
J.~M. Arrazola, A.~Delgado, B.~R. Bardhan, and S.~Lloyd, ``Quantum-inspired
  algorithms in practice,'' {\em arXiv preprint arXiv:1905.10415}, 2019.

\bibitem{realpe2017modeling}
J.~Realpe-G{\'o}mez, ``Modeling observers as physical systems representing the
  world from within: Quantum theory as a physical and self-referential theory
  of inference,'' {\em arXiv preprint arXiv:1705.04307}, 2017.

\bibitem{realpe2018cognitive}
J.~Realpe-G\'omez, ``Cognitive modeling of embedded observers can help make
  sense of quantum theory,'' {\em To appear}, 2019.

\bibitem{nielsen2002quantum}
M.~Nielsen and I.~Chuang, {\em Quantum Computation and Quantum Information}.
\newblock Cambridge Series on Information and the Natural Sciences, Cambridge
  University Press, 2000.

\bibitem{crutchfield1997statistical}
J.~P. Crutchfield and D.~P. Feldman, ``Statistical complexity of simple
  one-dimensional spin systems,'' {\em Physical Review E}, vol.~55, no.~2,
  p.~R1239, 1997.

\bibitem{realpe2019can}
J.~Realpe-G\'omez, ``Can cognitive science help us understand quantum
  theory?,'' {\em {\em To appear in} Journal of Cognitive Science}, 2019.

\bibitem{mezard2009information}
M.~Mezard and A.~Montanari, {\em Information, physics, and computation}.
\newblock Oxford University Press, 2009.

\end{thebibliography}
\end{document}